%% file: main.tex
\documentclass[sigconf]{acmart}
\usepackage[english]{babel}
\usepackage{amsmath}
\usepackage{tikz}
\usepackage{amsthm}
\usepackage{hhline} % used for the top-most horizontal line! in order to fix the top left corner cell empty
\usepackage{xcolor}
\usepackage{romannum}
\usepackage{graphicx}
\usepackage{caption}
\usepackage{subcaption}
\usepackage{paralist}
\usepackage[ruled,linesnumbered,vlined]{algorithm2e}
\usepackage{color,colortbl}
\usepackage{url}
\usepackage{amsfonts}
\usepackage{booktabs}
\usepackage[OT1]{fontenc}
\newcommand*{\rom}[1]{\expandafter\@slowromancap\romannumeral #1@}

\newcommand\sys{Bow-Tie }

\SetCommentSty{mycommfont}

\AtBeginDocument{%
  \providecommand\BibTeX{{%
    \normalfont B\kern-0.5em{\scshape i\kern-0.25em b}\kern-0.8em\TeX}}}

\setcopyright{none}
\copyrightyear{}
\acmYear{}
\acmDOI{}

\acmConference[Preprint '22]{}{ , 2022}{, }
\acmPrice{}
\acmISBN{}

\begin{document}
\title{Stopping Silent Sneaks: Defending against Malicious Mixes with Topological Engineering}

\author{Xinshu Ma}
\affiliation{%
  \institution{University of Edinburgh}
  \streetaddress{}
  \city{}
  \country{}}
\email{x.ma@ed.ac.uk}

\author{Florentin Rochet}
\affiliation{%
  \institution{University of Namur}
  \streetaddress{}
  \city{}
  \country{}}
\email{florentin.rochet@unamur.be}

\author{Tariq Elahi}
\affiliation{%
  \institution{University of Edinburgh}
  \streetaddress{}
  \city{}
  \country{}}
\email{t.elahi@ed.ac.uk}

\renewcommand{\shortauthors}{Ma, and et al.}

\begin{abstract}
Mixnets provide strong meta-data privacy and recent academic research and industrial projects have 
made strides in making them more secure, performance, and scalable. In this paper, we focus our work 
on stratified Mixnets---a popular design with real-world adoption---and identify that there still exist 
heretofore inadequately explored practical aspects such as: relay sampling and topology placement, 
network churn, and risks due to real-world usage patterns. We show that, due to the lack of incorporating 
these aspects, Mixnets of this type are far more susceptible to user deanonymization than expected. In 
order to reason and resolve these issues, we model Mixnets as a three-stage 
``Sample-Placement-Forward'' pipeline, and using the results of our evaluation propose a novel Mixnet 
design, Bow-Tie. Bow-Tie mitigates user deanonymization through a novel adaption of Tor's guard 
design with an engineered guard layer and client guard-logic for stratified mixnets. We show that 
Bow-Tie has significantly higher user anonymity in the dynamic setting, where the Mixnet is used over a 
period of time, and is no worse in the static setting, where the user only sends a single message. We 
show the necessity of both the guard layer and client guard-logic in tandem as well as their individual 
effect when incorporated into other reference designs. We develop and implement two tools,  1) a mixnet 
topology generator (Mixnet-Topology-Generator (MTG)) and 2) a path simulator and security evaluator  
(routesim) that takes into account temporal dynamics and user behavior, to assist our analysis and 
empirical data collection. These tools are designed to help Mixnet designers assess the security and 
performance impact of their design decisions. Ultimately, \sys is a significant step towards addressing the 
gap between the design of Mixnets and practical deployment and wider adoption because it directly 
addresses real-world user and Mixnet operator concerns.
	
%Mixnets are a fundamental type of anonymous communication system and recent
%academic research has made progress in designing Mixnets that are scalable,
%have sustainable communication/computation overhead, and/or provable security.
%We focus our work on stratified Mixnets, a popular design with real-world adoption. 
%The security of many designs rely on the \textit{anytrust} assumption
%where at least one server in the user's path must be honest.
%We identify the critical role Mixnet topological configuration algorithms play for user anonymity, and
%propose Bow-Tie, a performant topological engineering design for Mixnets that further ensures the 
%anytrust assumption 
%holds realized by introducing guard mixes.
%To draw actionable conclusions,
%we perform an analysis of the best realistic and resource-bounded adversarial
%strategies against each of the studied algorithms, and evaluate security
%metrics against each best adversarial strategy. Moreover, we highlight the need
%for a temporal security analysis and develop \texttt{routesim}, a simulator to
%evaluate the effect of temporal dynamics and user behaviors over the Mixnet.
%The resulting security notions are complementary to the state-of-the-art
%entropic definitions. The simulator is designed to help Mixnets developers
%in assessing the devil in the details resulting from design decisions.
%Ultimately, our results suggest strong potential improvements to current designs and
%guidance for shaping Mix networks.
\end{abstract}




\maketitle

\input{body}

\bibliographystyle{ACM-Reference-Format}
\bibliography{ref}

\appendix
\input{8_appendix}

\end{document}

%% file: body.tex
\input{1_introduction.tex}
\input{2_motivations.tex}
\input{3_modeling.tex}

\input{4_topologies.tex}
\input{5_strategy}

\input{5_behaviour.tex}
\input{6_extension}
\input{6_related_work.tex}

\input{7_conclusion.tex}

\section*{Artifact}
\texttt{routesim} is a Rust software open-sourced for the reviewing process at
\url{https://anonymous.4open.science/r/routesim-4735/}. \texttt{MTG} and various python scripts are open-sourced at
\url{https://anonymous.4open.science/r/MixnetConstructionSimulator-FD0F/}

%% file: 1_introduction.tex
\section{Introduction}
% \begin{enumerate}
%     \item Mass surveillance and the importance of anonymous communication systems and their deployment (put into practice). The general theme of this paper is privacy infrastructure and how to deal with adversarial resources, specifically in Mix networks.
    
%     \item Tor is both deployed and popular but it is not secure against global adversaries. Conversely, Mixnets are secure against global adversaries are not popular or practically deployed due to many practical concerns.
    
%     \item recent interest in mix networks and their deployment (DC-net like,
%       Vuvuzela, Loopix). However, an overlooked part of this space is how an
%       adversary that is allowed to insert their resources into the network can
%       break them. How this work fits with that effort and what we bring to the
%       table. Some of these works (e.g., Atom) offer an anytrust assumption
%       within their mixing design to protect against Network insiders (we
%       assume the anytrust assumption does not hold; and we evaluate the issue
%       in practice).

%     \item We investigate network topology and path selection as a critical aspects of the security of a mix network populated by adversarial resources. 
% \end{enumerate}

Since the ``Five Eyes'' mass surveillance disclosures by Snowden high-lighted real-world adversaries' pervasive and global nature, we observe a
greater community focus on strong meta-data privacy to protect and improve communication
protocols on the Internet~\cite{rfc7258}. Tor, with $\approx 8$ million daily users~\cite{torusage},  provides limited protection against a global adversary and traffic analysis attacks~\cite{murdoch2005low, johnson2013users, rochet2018dropping}. Thus, there is a
resurgent interest in mix networks (Mixnets)~\cite{chaum1981untraceable}---once considered impractical to deploy---with many recent proposals from academia~\cite{van2015vuvuzela,
  kwon2017atom, kwonxrd, tyagi2017stadium, piotrowska2017loopix, chaum2017cmix,
  karaoke} and industry~\cite{diaz2021nym} that have strong security guarantees and improved
performance at scale.

% Since the NSA mass surveillance disclosure by Snowden, we observe a
% greater focus on privacy to mitigate such a threat and improve communication
% protocols on the Internet~\cite{rfc7258}.
% End-to-End Encryption (E2EE) to protect message confidentiality is
% gaining mainstream popularity, but users are still susceptible to \textit{meta-data}
% leaks that allow adversaries to link communication parties together. Tor, the
% most popular low-latency anonymity network with $\approx 8$ million daily users~\cite{torusage},
% protects user metadata privacy
% by routing their
% traffic through a network of outsourced relays, such that a local eavesdropper
% cannot link the input and the corresponding recipient, assuming the
% eavesdropper does not hold a position covering both ends of a Tor circuit.
% Indeed, given Tor's
% limited security guarantees against a global adversary and traffic analysis
% attacks~\cite{murdoch2005low, johnson2013users,
%   rochet2018dropping}, there is a
% resurgent interest in mix networks (Mixnets)~\cite{chaum1981untraceable}, by
% which many recent systems with stronger security guarantees as well as improved
% scalability have been proposed in both academia~\cite{van2015vuvuzela,
%   kwon2017atom, kwonxrd, tyagi2017stadium, piotrowska2017loopix, chaum2017cmix,
%   karaoke} and industry~\cite{diaz2021nym}.

The security of many known designs, such as Vuvuzela~\cite{van2015vuvuzela},
Karaoke~\cite{karaoke}, Loopix~\cite{piotrowska2017loopix}, and
Nym~\cite{nym2021} rely on the \textit{anytrust} assumption where at least one server in
the user's path must be honest. In other words,  security comes from distributing trust across many relay operators.
%is the oft 
%overlooked problem of how to \textit{select resources and place} them in the network to weaken the 
%anytrust assumption.
% >>>>>>> 5486d7ae7a7e8c8fc1b1712570e30be70747f90c
%Similar previous work has thus far been neither
%thorough nor easy to deploy, missing important characteristics of the network,
%such as that an adversary is able to introduce new mixes into the Mixnet.
% Why the problem is important
Practical real-world designs distribute trust \textit{and} provision network resources by drawing from 
third-parties, such as volunteers or for-profit participants, on which the
 network applies light (e.g., Tor's path selection IP restrictions) or no
constraints. These third-parties can be malicious and it is critical that the mixnet design resist their influence. 
% \tariq{NOT SURE IF THIS SHOULD BE LATER ON: In addition most networks use all of the resources available, however, there are scenarios---such as times of low user activity and traffic---where a fraction of all available resources provides higher security and performance; an insight that we take advantage of in our proposal.}
In Mixnet literature, the security
analysis typically considers active attacks like traffic analysis~\cite{agrawal2003measuring}, (n-1)~\cite{serjantov2002trickle},
% <<<<<<< HEAD
% and Denial-of-Service (DoS) attack~\cite{borisov2007denial}. However, taken for granted is the often overlooked problem of how to \textit{select resources and place} them in the network to strengthen the anytrust assumption.
% =======
and Denial-of-Service (DoS)~\cite{borisov2007denial}. In addition, users can also be deanonymised by passive adversaries whenever a message traverses a path composed entirely of adversarial relays. 

However, the literature typically takes for granted real-world issues such as \textit{network configuration and routing, network churn, and risk due to real-world usage patterns}. In this paper, we consider the impact of these practical concerns and investigate designs that strengthen the anytrust assumption while minimizing performance degradation. 
% In real-world settings, there are often scenarios where the demand, i.e. number of online users or traffic, is low. To investigate the effect on security and performance we also consider using a fraction of all available relays.
% , which is 
% relevant for scenarios where the user or traffic demand is low. number of users or quantity of traffic is low and distributing across all available 
% Mixnodes would have a detrimental effect on security and performance. 
%An adversary would populate as many nodes as possible into the
%network since all messages travel over the fully malicious paths will be
%deanonymized. 
%In this paper, we investigate how selection and placement can improve the security of the anytrust 
%assumption, while accounting for performance and network churn.
%We observe that the network could be built from further more
%attention to the anytrust assumption, yet accounting for current users' needs
%and network churn. 
% In this work, what we did and what we found--key takeaways, our findings.
We present the first thorough analysis of continuous-time stratified
Mixnet designs, 
%first proposed by Danezis~\cite{danezis2003mix}, 
and the implications on the security of typical users
against realistic resource-bounded strategic adversaries in the
network. 
% \tariq{IS THIS STAYING OR LEAVING?: Our \textit{snapshot} analysis of the state-of-the-art academic
% and industrial Mixnet approaches shows that an adversary that controls $30\%$ of the resources may on average deanonymize $42.9\%$
% of all traffic by merely observing and logging.} 
Our \textit{temporal} analysis, where we model an
adversary cumulatively deanonymizing users over time, shows that in the state-of-the-art reference designs close to $100\%$ users are expected to use a fully malicious route in 
about one week of email activity over the
Mixnet.
%This clearly shows the dramatic impact these
%malicious mixes can have on security.
Overall, the adversary is able to deanonymize a significant portion of network
traffic running a realistic amount of bandwidth and quantity of nodes. This
implies that the anytrust assumption is not easy to maintain in real Mixnet deployments. 
%In our design, we show how to mitigate the security impact of
%passive malicious mixes by incorporating guard design into Mixnet 
%topological construction.

%We propose \textit{Bow-Tie}, an efficient mixnet design that
%limits the adversary's power to deanonymize
%traffic via fully malicious paths from two aspects---i) topological construction
%and ii) message routing. 
%We identify the tradeoff
%between security and performance for \textit{Mixnet configuration} algorithms
%and we believe Bow-Tie finds an effective balance between mitigating the
%adversarial advantage and offering low message queuing latency.
%Furthermore, we introduce \textit{guard mixes}, inspired by guard relays in the Tor network, to mitigate the over-time client exposure. Our results show that 
%Bow-Tie provides a practical
%solution of accommodating guard design to stratified topology configuration and
%facilitates network churn in a reasonably realistic setting. 

%Furthermore, we examine
%how individual behaviors affect users
%anonymity. \tariq{Is this sentence about routesim and rendezvous, perhaps? I am not sure.}
%%, and our analysis shows that a rendezvous-based protocol might be a
%%better idea for Mixnet-based systems.

\noindent\textbf{Contributions.}
%\quad Our paper makes the following main contributions:

\begin{inparaenum}
  \item We propose Bow-Tie, a novel practical and efficient Mixnet design that
    mititgates the over-time client exposure to adversarial mixes and
    strengthen the anytrust assumption. We realize it by adapting and re-engineering
    the concept of guards from Tor~\cite{tordesign} to stratified Mixnets.
    %that defines
    %how to limit the security impact of malicious mixes with careful
    %topological constructions and path routing constraints realized with our Guard Design.

  \item We present
    an empirical security analysis of the stratified Mixnet against reasonably
    realistic adversaries from the metrics of i) fully-compromised
    traffic fraction and ii) time-to-first compromise. We show how these
    results relate to a newly deployed Mix network and how it could be
    significantly improved.

  \item We develop the
    \texttt{routesim} simulator, a tool that can calculate a user's
    expected deanonymization probability over time, given a configured
    network topology and communication patterns. \texttt{routesim} may be used to
    shed light on the security impact of various design choices, in Bow-tie and other designs as well.
  
  %\item MTG also?
  \end{inparaenum}

% The rest of the paper is organized as follows. We give the background
% behind our work in Section~\ref{sec:2_background}. In
% Section~\ref{sec:3_model}, we define the system model and define the realistic threat model.
%  % and define our design goals.
%  We introduce
% our Bow-Tie design including the Mixnet engineering and guard layers in
% Section~\ref{sec:4_topologies}. Then we evaluate the topological configuration
% approach in Section~\ref{sec:5_static_analysis} and analyze the users anonymity
% in Section~\ref{sec:5_behavioural}. We give related work in
% Section~\ref{sec:related_work} and conclude with potential future work in
% Section~\ref{sec:conclusion}.
% \xinshu{introduce more description of highlights.}

%% file: 2_motivations.tex
\section{Background and Motivation}
\label{sec:2_background}
Mixnets are a fundamental type of anonymous communication system, 
composed of a set of Mixnodes that provide sender and sender-recipient anonymity by reordering 
messages in addition to transforming them cryptographically, enabling
message untraceability. 

Unfortunately, early Mixnets have practical disadvantages such as high latency, poor scalability, 
and high-performance overhead that hinder their real-world deployments.
Recent academic research~\cite{van2015vuvuzela,tyagi2017stadium,kwon2017atom,
	chaum2017cmix,piotrowska2017loopix,piotrowska2017loopix,karaoke} has made 
progress in designing Mixnets for anonymous communications with 
developed scalability and sustainable communication/computation overhead, or
provable security. These developments have found their way into the industry, with the
foundation of a startup company---Nym~\cite{diaz2021nym}, whose goal is to create a sustainable 
anonymous communication network based on the Loopix continuous-time mix 
design~\cite{piotrowska2017loopix} through monetary incentive schemes.

\noindent\textbf{Network topology.}\quad Mixnets can be arranged in many topologies. Mesh, 
cascade, and stratified are some of the most common. In this paper, we focus on the stratified topology~\cite{danezis2003mix} 
due to the evidence that it is both as, or more, secure and performant as the other two~\cite{dingledine2004synchronous, diaz2010impact}.
In a stratified topology, the network is constructed from several `layers'.
Each Mixnode is placed in a single layer, and each layer can only communicate
with the previous and next ones. Generally, layers are equally sized for
performance reasons, although this is not a strict requirement.  
%Let $n$, a parameter targeting the
%number of layers in the network, selected by the designer or the operator of
%the Mixnet.  The layers are connected such that all Mixnodes in layer $i$ can
%only forward messages to Mixnodes in layer $i+1$.  
At the last layer, the
messages are delivered to their intended destination (or wherever the user's
inbox is hosted).

% \noindent\textbf{Mixnode allocation.}\quad While most designs assume that all
% available Mixnodes are used in the network, this is not a strict requirement.
% Good reasons could leave some relays unused, due to an overprovision of
% capacity and its impact on the Mixnet's mixing quality. The Nym network is a
% real-world example where not all available Mixnodes are used to ensure good
% mixin.

\noindent\textbf{Path selection/routing.}\quad Messages are forwarded through a
Mixnet by going through a Mixnode in each layer. This multi-hop path through
the network provides the sender and sender-recipient anonymity property. It is
therefore critical that the route through the network is not biased or
otherwise manipulated by an adversary. Most Mixnet designs route messages by
`bandwidth weight'. That is, the probability of selecting a Mixnode in
layer $i+1$  is proportional to the proportion of its bandwidth to the sum of
all Mixnodes bandwidths in that layer. An alternative is to route packets by
choosing uniformly at random. We experiment with both approaches in this work
and show that uniform selection is inadequate for performance
and can be marginally better or worse from a security perspective, depending on
the adversary resource endowment.

\noindent\textbf{Continuous-time mixing.}\quad Various mixing strategies have
been proposed in the literature. Timed, threshold, pool, and continuous-time
are the main types.  We focus on continuous-time mixing in this paper since it
has emerged as a good trade-off between security and performance. In
continuous-time mixing, each message is independently delayed at each mix on
its path. This delay is selected by the sender of the message. To offer some
level of security against timing attacks, the delay is drawn from an
exponential distribution because of its memoryless property (i.e., observing a
message going out does not give information about when other messages are
scheduled to go). Indeed, Loopix, and by extension Nym, use this mixing
strategy, providing real-world relevance.

% \noindent\textbf{Full path compromise.}\quad An adversary may compromise the
% anonymity of a message, i.e., discover the sender and recipient, whenever that
% message traverses a path composed entirely of Mixnodes controlled by the
% adversary. To measure the likelihood of this occurrence, we count the number of
% possible fully compromised paths through the network in proportion to
% the total number of paths through the network. This passive attack is
% impossible to detect and not possible to know when and if a message has been
% fully compromised. Active attacks can improve upon this, however, they
% are also more detectable and hence riskier for the adversary to mount. In this
% paper we focus our attention to passive fully compromise as a relative
% measure of the designs we will consider.

\noindent\textbf{Anytrust assumption.}\quad Many of these systems, Loopix
included, rely on the anytrust assumption: as long as there is one
honest Mixnode in a path, then the user's message cannot be fully compromised.
However, we show that this assumption breaks quickly, and for every users, as
soon as one considers temporal aspects in the Mixnet usage. Our work considers
this problem when designing Mixnet topologies, and as a consequence,
significantly strengthen how realistic this assumption is for the users.

%% file: 3_modeling.tex
\section{Threat Model}
\label{sec:threat_model}

In general, we consider the adversary who can observe all internal states of
controlled mixes and can locally drop, inject or delay traffic. A global active
attacker may be able to compromised more paths using (n-1)
attacks~\cite{serjantov2002trickle}, or Denial-of-Service (DoS)
attacks~\cite{borisov2007denial}, this behaviour is also more detectable and
risks ejection from the network. It is therefore realistic to consider a more
subtle and discrete adversary who traces the messages that travel through paths
composed entirely of their malicious Mixnodes. 

\noindent\textbf{Adversary Resources.}\quad At an abstract level, we assume the adversary has a certain fixed amount of 
network resources at their disposal. This could be in the form of bandwidth,
relays, financial assets, or some 
other scarce resource like reputation. Note that it is possible to swap one criterion  with the other, since their function is the same; limiting the adversary to control only a fraction of the paths through the network. 
When deciding on its resource allocation, we allow the adversary to take advantage of the network configuration
or path selection algorithms to influence a network distribution that maximizes their presence on user paths. 
This passive attack is impossible to detect. 
%Active attacks can improve upon   
%this, however, they are also more detectable and hence riskier for the adversary to mount.

\noindent\textbf{Adversary Goals.} \quad
% <<<<<<< overleaf-2022-06-28-2154
The adversary's aim is to maximise end-to-end path compromise rates by causing the mixnet configuration step to optimally place the malicious mixes into the mixnet layers in a way that maximizes their ability to passively deanonymise users.

%% file: 4_topologies.tex
\section{\sys Design}
\label{sec:4_topologies}
\begin{figure}[!t]
  \centering
  \includegraphics[width=.30\textwidth]{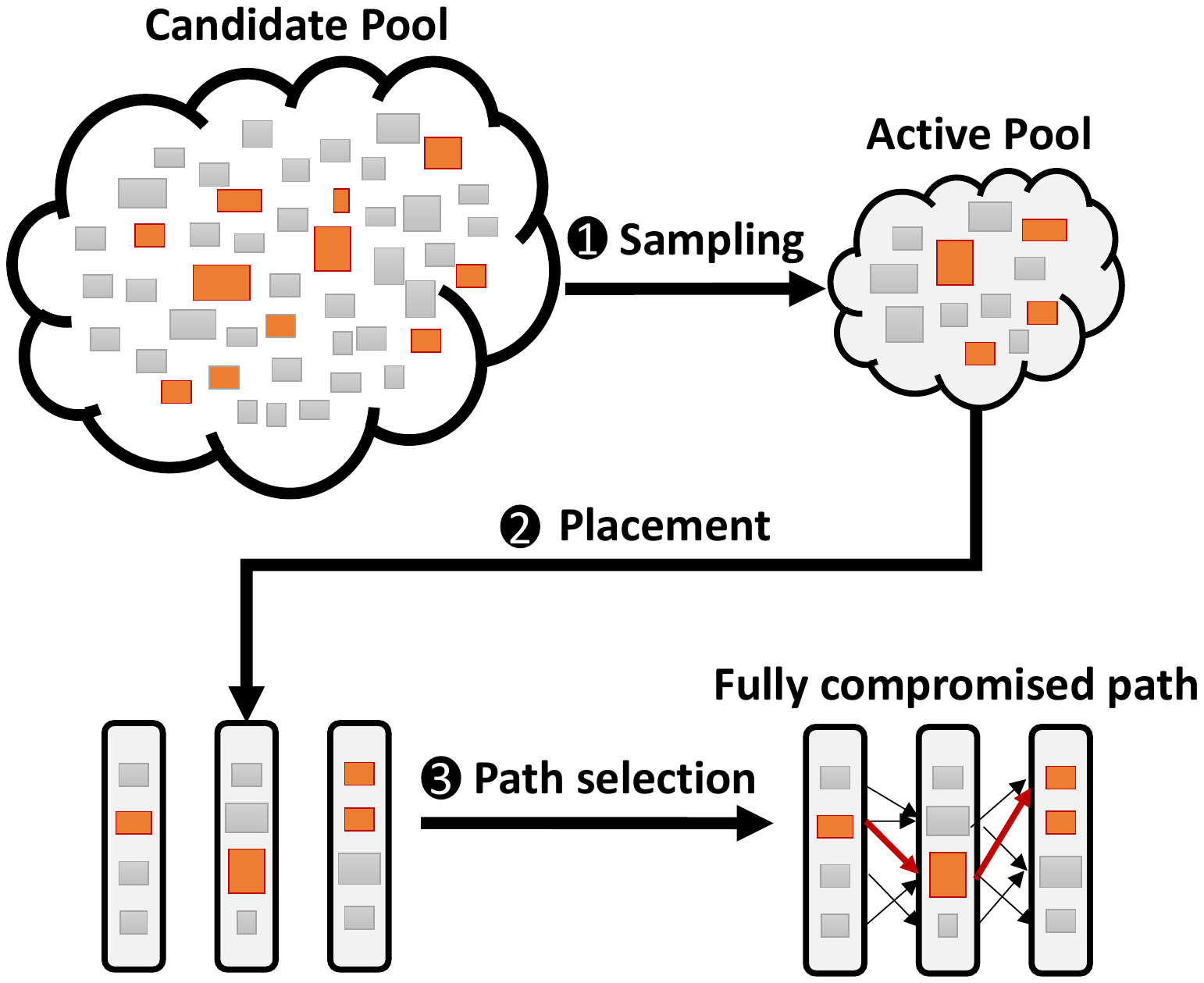}
  \caption{Three-steps basic pipeline when configuring Bow-Tie.}
  \label{fig:system_model}
\end{figure}

We propose a new design, Bow-Tie, a three-step pipeline
(Figure~\ref{fig:system_model}) to configure and use the stratified Mixnet. In Bow-Tie,
we discretize time into periodic epochs, where the network can be reconfigured
and clients updated with the latest topology.  Furthermore, to reflect
real-world resource availability, Mixnode bandwidths are heterogeneous and the
Mixnet is tunable in order to adjust the network size to suit the
volume of incoming traffic. 
We assume an honest-but-curious
\emph{Configuration Server} (CS) that periodically (re)configures the network,
ensuring node sampling and placement is correct.
% (Figure~\ref{fig:system_model}). 
It is common for anonymous communication
systems to depend on a trusted party for efficiency reasons, such as Directory
Authorities (DA) in Tor. Note that a malicious CS might collude with the
adversary and enable malicious Mixnodes to be sampled and placed in the network. 
However, we believe that it 
is a reasonable assumption in practice because even with the collusion,
% since 
the CS's deviations from honest behavior will eventually be 
detected 
by all participants over time, since the probability distribution of node selection will deviate from 
expectation. We aim to remove the honest-but-curious CS assumption as future work.

%Tor defends against this threat with a voting protocol that the DAs participate in when generating the 
%latest network documents. 

\subsection{Bow-Tie Characteristics}

Bow-Tie builds Mixnet topologies
around the following important criteria.
%For each epoch, the \textit{Configuration Server} (CS) periodically
%re-configures the \sys Mixnet.  While \sys is amenable to any number of
%layers, we will discuss a 3-layer network as it provides a good trade-off
%between security and performance\cite{guirat-wpes}. 

\noindent\textbf{Mitigating Client Enumeration.}\quad
In general, users will fall victim to full path compromise the more (or longer)
they use the system. Eventually, all users will have at least one of their
messages traverse a fully compromised path (see
Section~\ref{sec:5_behavioural}). This problem is also referred to as client
enumeration, where the adversary observes at least one message from every
single user of the system.  One successful strategy is to limit the exposure of
clients to all nodes in the network by restricting the paths clients select.
Tor realizes this strategy with its \textit{Guard Design}, which has undergone
several refinements since its initial
proposal~\cite{wright2004predecessor, prop271, prop310, rochet-claps}---that
\begin{inparaenum} 
    \item Ensures quality of service, and
    \item Limits the size of the set of guards the client is exposed to.
\end{inparaenum}
Bow-tie introduces a novel guard design that uses restricted topologies and
client-side logic specifically targeting mixnet integration. The design is
supported by empirical results (in Section~\ref{sec:5_behavioural}).

Unlike Tor, where guards must be placed in the first layer, in 
Mixnets we have more freedom to choose in which layer to place guards. 
For a client building routes of length $L$ within a Mixnet,
there is a subtle Performance-Security trade-off in choosing either the first node as
the guard (more performance) versus choosing one of the middle positions (more
secure in specific settings). \sys adopts guards in the middle position, and
the security metrics we explore in Section~\ref{sec:5_behavioural} are
independent of the position choice.  Appendix~\ref{app:choosing_position} covers a
discussion shedding light on the subtleties.

\noindent\textbf{Accommodating Network Churn.}\quad
\label{subsec:churn}
Another fundamental issue in real-world mixnet deployments is Network churn,
which is a typical and natural phenomenon in volunteer-resourced networks. One
of its effects is to increase the clients' exposure to potentially malicious
guards~\cite{elahi2012changing}. Similar to Tor, the client is required to
prefer using an older guard---until they go offline---before touching a new
guard. Thus, the more unstable the guards are, the more guards a client will
touch, which implies a higher risk of choosing a malicious guard. Therefore,
putting the most stable Mixnodes into the guard layer ensures that the guard
list of each client grows at a slower pace.

\noindent\textbf{Good Performance \& Low Cost}\quad
Bow-Tie also considers the performance of generated network topologies. In the stratified Mixnet, the transmission bottleneck comes from a layer with the minimum bandwidth. To mitigate this, we model the placement of Mixnodes into a Bin-packing problem, which improves the network performance by constrainting that the total bandwidth of each layer is approximately equal. 
% network performance of the generated topologies with regard to avoid the 
The results show that Bow-Tie strikes a good balance between anonymity and performance (in Section~\ref{subsubsec:performance}) and runs in an efficient manner.

\subsection{Bow-Tie Detail Description} 
\label{subsec:design_overview}

\begin{figure*}[t!]
	\centering
	\includegraphics[width=.95\textwidth]{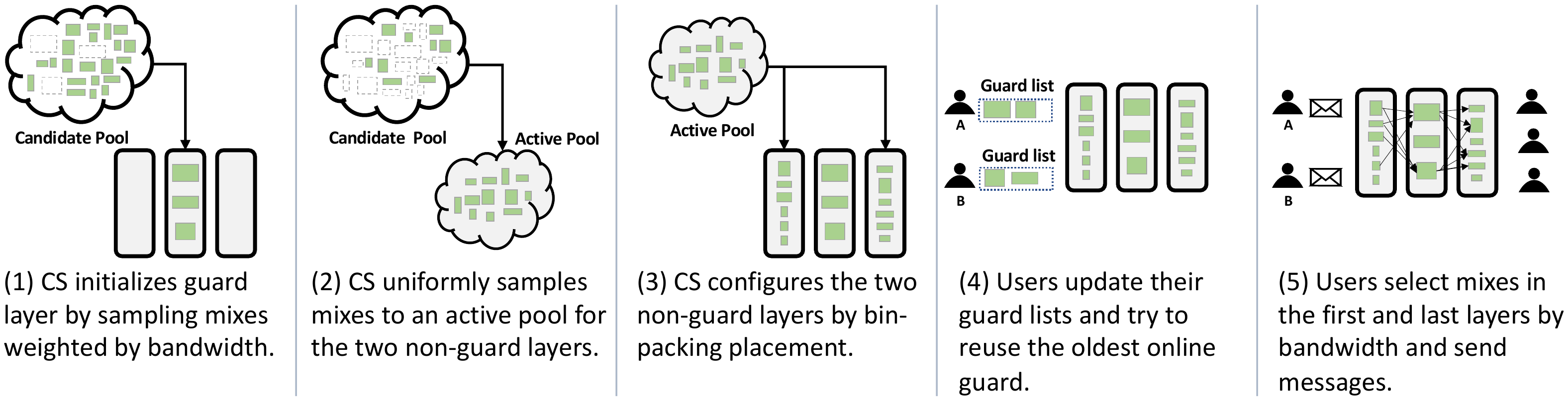}
	\caption{High-level overview of \sys for Mixnet initialization and message
		routing. Green rectangles indicate that malicious mixes are indistinguishable.}
	\label{fig:BT_overview}
\end{figure*}

Steps to create and maintain a Bow-Tie Mixnet are depicted in Figure~\ref{fig:BT_overview} and detailed next.

\subsubsection{Mixnet Initialization}\label{subsec:mixnet_init} The bandwidth of the candidate pool, $P_{bw}$, is the sum of all 
the available relays' bandwidths. A predetermined sampling fraction, $h$, of $P_{bw}$ is the total 
bandwidth of the active pool from which the generated Mixnet is populated. Each layer accounts for  
$\frac{1}{l} 
\times h$ of $P_{bw}$ in a l-layer Mixnet. We consider the case $l=3$.

%The total bandwidth of the nodes in the generated Mixnet is a
%predetermined fraction, $h$ (sampling fraction, see (1) in Figure~\ref{fig:system_model}), of $P_{bw}$, 
%the bandwidth of the candidate pool,
%and each layer accounts for $\frac{1}{l} \times h$ of $P_{bw}$ in a l-layer Mixnet.

% \xinshu{configure all layers for epoch0}
% connfigure all layers for epoch0

\begin{inparaenum} 
  \item \emph{Initialize the Guard Layer.}
The CS initializes the guard layer in the first epoch, $i=0$, by sampling a total of
	$\frac{1}{3}{h\times P_{bw}}$ weighted by bandwidth from the candidate
	pool. The rational is to ensure that the guard layer has 
%	obtain a total set of guards bandwidth of
	$1/3$ of the overall active network bandwidth with the remaining to be distributed evenly across the 
	remaining 2, i.e. $l-1$, layers.
%	, since all guards are in the same layer.
% in conjunction with or as replacements.
% , with $P_{bw}$ the bandwidth of the candidate pool, and $h$ the sampling fraction.

  \item \emph{Initialize the Guard Set.} 
% \textit{Guard Layer and Sets Initialization.}\quad 
% These Mixnodes are placed in $AG$. 
The guard set $G$ consists of three subsets: \textit{Active Guard (AG)}, 
\textit{Backup Guard (BG)}, and \textit{Down Guard (DG)}. All nodes in the 
initialized guard layer are elements of $AG$. 
The CS then samples an additional tolerance fraction $\tau$\footnote{The value of $\tau$ is defined as $\tau = c\times$ churn rate. In this paper, we set $c = 1$.} from
	the candidate pool by bandwidth as $BG$. $DG$ is empty at this stage. The rational behind these sets 
	is to minimize client exposure by remembering which nodes were used as guards, even if they go 
	offline for a period. That is, clients can revert to a previously used but offline guard whenever it appears 
	online again.

% CS sample a slightly larger-than-required total guard bandwidth from the candidate 
% pool and form the guard set $G$. After updating each mix's online status, the
% CS forms the guard layer, selecting nodes from $G$ to form the \textit{Active
%   Guard (AG)} set. The remaining guards of $G$ compose
% the \textit{Backup Guard (BG)} set, and the \textit{Down Guard (DG)} set for
% any offline guard. The CS will maintain $AG$ within an appropriate
% size and bandwidth, preferring returning guards that 
% were offline and picking from the Backup set instead of sampling new ones from
% the candidate pool. Finally, the CS may evict unstable guards, and if
% the $G$ bandwidth falls below the larger-than-required total guard
% bandwidth threshold, the CS introduces fresh guards from the candidate pool, as detailed in 
% Algorithm~\ref{alg:config_guard}.

  \item \emph{Initialize non-Guard Layers.}
Next, the CS uniformly samples a total of $\frac{2}{3} {h\times 
P_{bw}}$ Mixnodes from the candidate pool and places them using a bin-packing approach 
with the constraint that each layer has a similar amount of bandwidth. We convert the placement problem to 
\textit{one-dimensional bin packing problem
  (1BPP)}~\cite{scheithauer2017introduction}, where the objective is to pack
all items into a minimum number of bins while the total size of any bin is not
larger than the given capacity $c$. Thus the capacity is set to be slightly
larger ($\epsilon$ in Algorithm~\ref{alg:config_ordinary}) than the 
$\frac{1}{3} {h\times P_{bw}}$
bandwidth for each layer as it is difficult to aggregate several indivisible
entities to a precise cumulated bandwidth value.
  % 2/3 of the
% projected total active network bandwidth with uniform probability. 
% Then the CS
% adopts a Bin Packing approach to place Mixnodes into layers, constraining each
% layer to have a similar amount of bandwidth as $L_g$, as detailed in
% Algorithm~\ref{alg:config_ordinary}. 

% \textit{Binpacking Placement.}\quad We model the placement problem into a bin
% packing problem: let $m$ items (Mixnodes), each having a weight (bandwidth)
% $b_i$, $i\in{1,\cdots, m}$, and a number $l$ of identical bins (layers) of
% unlimited capacity. Then the objective is to pack all items into bins so that
% the total weight of each bin is approximately equal, i.e., the difference
% between the weights of the bins is minimal. Specifically, we convert to the
% \textit{one-dimensional bin packing problem
%   (1BPP)}~\cite{scheithauer2017introduction}, where the objective is to pack
% all items into a minimum number of bins while the total size of any bin is not
% larger than the given capacity $c$. Thus the capacity is set to be slightly
% larger ($\epsilon$ in Algorithm~\ref{alg:config_ordinary}) than the projected
% bandwidth for each layer as it is difficult to aggregate several indivisible
% entities to a precise cumulated bandwidth value.  We show in
% Section~\ref{subsubsec:compromised traffic} and
% Section~\ref{subsubsec:performance_analysis} that the Mixnet with an even
% amount of bandwidth in each layer limits, in the worst case, the adversary's
% ability to compromise traffic and allows higher network throughput as well.

\end{inparaenum}

% \textbf{Initialize the Guard Layer.} For the first epoch,
% the CS populates the guard layer by sampling Mixnodes weighted by their
% bandwidth. For subsequent epochs, the CS maintains the guard layer by keeping as many
% old and stable guard nodes in the guard layer as possible ((1)in
% Figure~\ref{fig:BT_overview}). As we will describe below, keeping the guard layer stable reduces client 
% exposure to malicious Mixnodes.

% \textbf{Mixnet Layer Configuration Details}\quad
% \label{subsec:configuration}
% We now describe how the guard and non-guard layers of \sys are created and
% maintained. 

% \noindent\textbf{Details.}
% Configuring the guard layer has three requirements: 1) limit the adversary's ability to enter it, and 2) maintain overall 
% network performance, and 3) mitigate the effects of network churn.
%To limit the adversary's ability to enter into the Guard Layer after its first
%initialization and maintain the overall performance due to the network churn,

\begin{figure}[t]
  \centering
  \includegraphics[width=.3\textwidth]{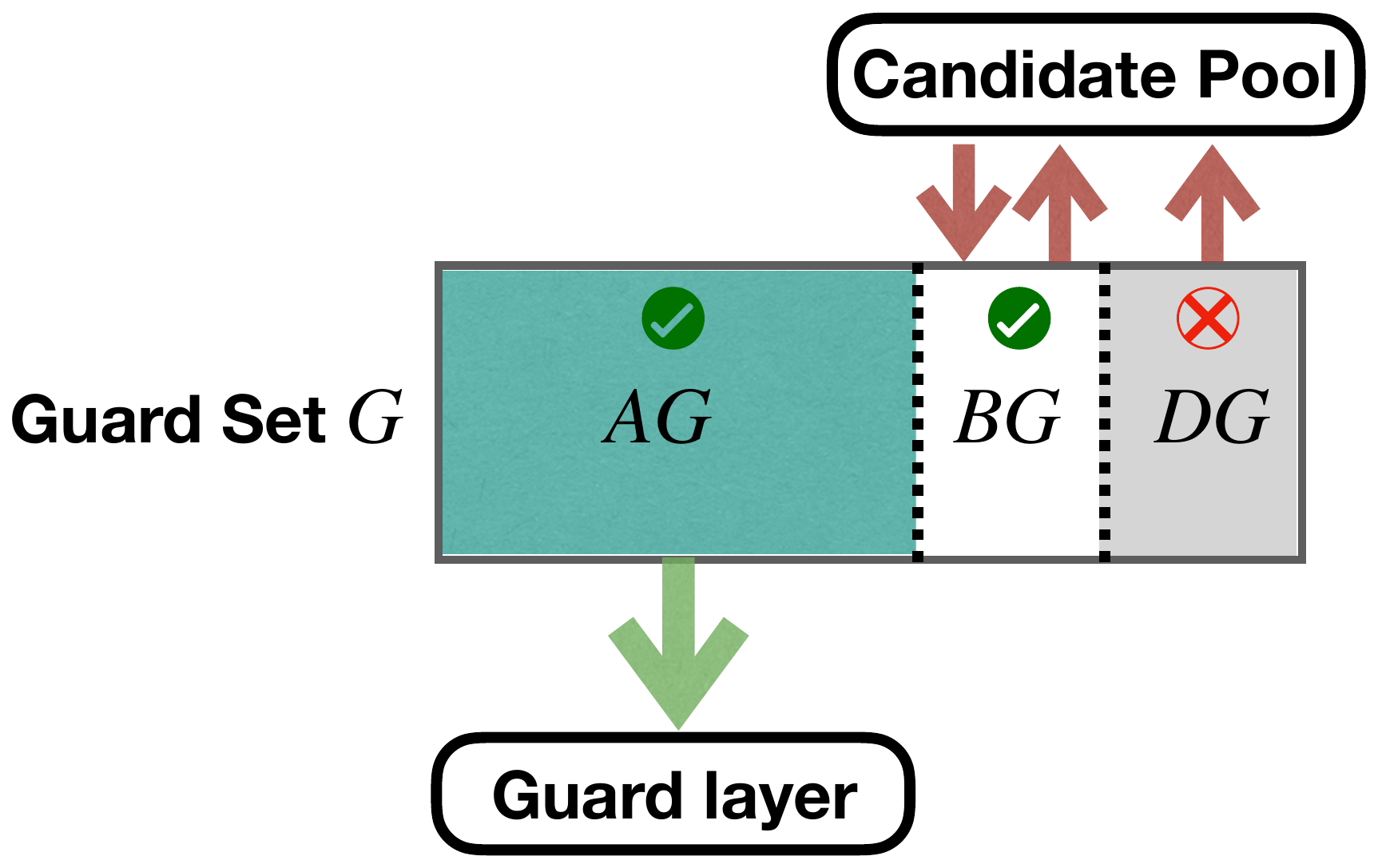}
  \caption{Guard Set composition and interactions.
%   At the end of each epoch, offline guards
%   are moved to $DG$ and CS checks. If the quantity of remaining online guards is insufficient, 
%   CS introduces fresh guards from candidate pool. If the quantity exceed the maximum threshold,
%   CS moves unstable guards to the candidate pool. After rearranging subset $AG$ and 
%   $BG$, all Mixnodes in AG are placed into the guard layer.
  }
  \label{fig:guard_maintain}
\end{figure}

% \textbf{Configuring non-Guard Layers.} At each epoch, the two remaining layers 
% are 
% reconfigured by the CS by uniformly sampling a set number of Mixnodes from the candidate pool and 
% placing them in a layer at random((2-3) in Figure~\ref{fig:BT_overview}).

% \noindent\textbf{Details.}
% For the remaining two layers, the CS samples from the candidate pool 2/3 of the
% projected total active network bandwidth with uniform probability. Then the CS
% adopts a Bin Packing approach to place Mixnodes into layers, constraining each
% layer to have a similar amount of bandwidth as $L_g$, as detailed in
% Algorithm~\ref{alg:config_ordinary}. 

\subsubsection{Mixnet Maintenance}
\label{subsec:mixnet maintenance}
\begin{inparaenum}
Clients need to learn about new Mixnodes, and offline Mixnodes need to be be
removed from the active pool. Moreover, to give a chance to nodes
remaining in the candidate pool to actually contribute, at the end of each
epoch, a new placement step is executed for non-guards layers. Finally,
specific maintenance for the guard layer is required at each epoch.

\item \emph{Guard Set Maintenance.}
For subsequent epochs $i >0$, the CS checks 
online/offline status of all nodes in $G$  and updates their stability information.
Offline nodes within $AG$ and $BG$ are moved to $DG$ the rest remain where they were.

In addition, CS checks if the bandwidth of $AG\cup BG$ is within the minimum threshold $T_{low}$ and 
maximum threshold $T_{high}$. 
In the case it is greater than $T_{high}$, nodes in $BG$ that have never have been selected into the 
guard layer
are dropped according to the ascending order of bandwidth$\times$stability\footnote{The stability values 
of nodes are evaluated by WMTBF and normalized to $0-1$ scale.}. This continues until the bandwidth is 
$\le$ 
$T_{high}$ or all eligible nodes in $BG$ have been evicted. In the case of bandwidth being lower than
$T_{low}$, the CS introduces fresh nodes from the remaining candidate pool by 
bandwidth$\times$stability to $BG$. Note that the CS tracks the online/offline status of each node and 
obtains their 
stability values; this scheme is detailed later.
By introducing new 
guard nodes or dropping unstable and slow ones, the CS maintains the whole guard set 
with high stability and sufficient capacity (Figure~\ref{fig:guard_maintain}). Please 
refer to Algorithm~\ref{alg:config_guard} in Appendix~\ref{app:algorithms}.

  \item \emph{Guard Layer Maintenance.}
Once the $G$ set is updated, the new
$AG$ set is generated by inheriting online guards from the old $AG$ and 
guards back online from the previous epoch's $DG$ set. To minimize the number
of guards users are exposed to, the CS records the number of epochs of active
operation as $t_{AG}$, for all Mixnodes in $G$, and selects the most stable ones based
on their WMTBF value. If $AG$ still does not meet the minimum bandwidth threshold, the CS samples 
some nodes from $BG$ by bandwidth$\times$stability to $AG$. In the end, all nodes in $AG$ are placed 
into the guard layer.

% Once the $G$ set is updated, the CS checks if bandwidth of current $AG$ meets the minimum threshold $T_{
% low}$. If the condition is satisfied, the CS simply place all nodes in $AG$ into guard layer. Otherwise, the CS 
% will sample nodes from $BG$ by bandwidth$\times$stability until it meets the minimum bandwidth 
% threshold. Similarly, all nodes in $AG$ are placed into guard layer.

  \item \emph{Non-guard layers Maintenance.}
The non-guard layers are refreshed through the same procedure as initialization. Please refer to 
Section~\ref{subsec:mixnet_init}.

  \item \emph{Stability Tracking.} To track the stability of each Mixnode, we use the metrics
Weighted Mean Time Between Failure (WMTBF)
 as also used by Tor~\cite{Tordir}. Briefly, online/offline states are
represented by $1/-1$ respectively in a discretized time interval. The weights
of these values are adjusted 
in proportion to their age from the current epoch. The rational is to discount epochs' values in proportion 
to their age such that very old epoch values would not significantly influence the WMTBF result. 

%in practice, note that careful consideration should be
%put on monitoring and detecting instability due to Denial-of-Service attacks
%before introducing new guards. Denial-of-Services could be attempts from the
%adversary to game the CS and get more nodes in. If the network is under Denial
%of Service, the CS should refrain from introducing new nodes.
\end{inparaenum}

 \subsubsection{Mixnet Routing}
 Once the network has been constructed, the Mixnet is ready for use.

\begin{inparaenum}
  \item \emph{Client-side guard logic.} 
When a user first uses the Mixnet they sample a defined number of Mixnodes, proportional to
their bandwidth, belonging to the guard layer and adds these to
their guard list. The user's guard list will grow over time. To limit
its growth and reduce the user's exposure to malicious Mixnodes in the guard
layer, the user's client only adds a new guard if all existing guards in the list
are offline. Whenever a new path is required, the client tries to reuse
guards from oldest to most recent ((4) in Figure~\ref{fig:BT_overview}).

  \item \emph{Path selection.} 
With an online guard chosen, the user selects Mixnodes in the first and last
layer weighted by bandwidth and sends the message through this fresh route 
((5) in Figure~\ref{fig:BT_overview}).
% For each of their messages, clients use 
% Bandwidth-weighted path
% selection, in which Mixnodes are selected from each layer with a
% probability proportional to their bandwidth, to construct paths to their destinations.
\sys uses Bandwidth-weighted path selection since it has better performance and security than the 
alternative, random path selection, which does not help in protecting clients and incurs a significant 
performance cost (for details refer to Appendix~\ref{sec:random path selection}).

%An alternative is random path selection, in which the Mixnodes are selected from each layer with 
%uniform probability, however, this
%% selecting a route uniformly
%% at random 
%can delay, or marginally hasten attack success
%depending on the adversarial bandwidth distribution. Random routing does not help protecting clients 
%and it incurs a significant performance cost. For more detail, please refer to Appendix~\ref{sec:random 
%path selection}.
% our test results~\ref{sec:random path selection} show that the security
% and performance are worse compared to bandwidth path selection.
% Figure~\ref{fig:time} and Section~\ref{sec:5_static_analysis}.
\end{inparaenum}

 % \xinshu{Client side guard logic and path selection}
 % \xinshu{Why we use the bw path selection? The detail of random path selection are in appendix. Where does the bandwidth come from?}
% \textbf{Client Guard State.} 

\subsection{Bandwidth Discussion}
In our design, we use bandwidth as the sampling criterion since it is well 
established in the literature and in real-world deployments~\cite{Tordir}. However, 
other attributes can also be used~\cite{johnson2015avoiding,jaggard201520,jaggard2014representing}. 
As for how the bandwidths of the nodes are determined, the simplest way would be for
the nodes to advertise their bandwidth, however in such case, adversarial nodes could advertise
false information. There exist previously proposed bandwidth measurement 
systems~\cite{traudt2021flashflow,andre2018smartor,johnson2017peerflow,jansen2021accuracy,snader2009eigenspeed}
 have improved the security, accuracy, and efficiency of estimating capacity in Tor network. A similar 
approach may be adopted here, however the exact solution to this problem is out of the scope for this 
present paper.

%% file: 5_strategy.tex
\section{Methodology}
\label{sec:evaluation methodology}
We now describe the security metrics, reference algorithms, and adversary model used in our evaluation. Since we consider a realistically constrained and strategic adversary, we identify the optimal adversarial resource allocation strategy that will be employed in our evaluations. 

%We present security metrics for the threat of client enumeration in mixnets. We consider a realistic and intelligent adversary that has the resources to run Mixnodes in mixnets. Specifically, we take bandwidth as the limiting resource and consider an adversary that allocates malicious bandwidth to Mixnodes.

\subsection{Security Metrics}
\label{subsec:security metrics}
We wish to evaluate how well a mixnet design is able to resist compromised
mixes (as defined in Section~\ref{sec:threat_model}). We use the following metrics: 
\begin{itemize} 
	\item \emph{Time to first compromise}: The expected time it takes until a user
			has their first message 
    traverse a fully compromised path. This is a dynamic metric since it is affected by usage patterns and is 
    useful to reason about user behavior.
    \item \emph{Compromised fraction of paths}: The expected fraction of total paths in the network 
    topology that are fully compromised (i.e. composed entirely of the adversarial relays). This is a static 
    metric since it is not affected by usage patterns. 
    \item \emph{Guessing entropy}: 
%    The expected number of Mixnodes the adversary needs to compromise in order to compromise a 
%particular message. 
    We also consider an \textit{active external} adversary for the scenario where she targets a specific 
    message 
    sent from a particular user. We wonder how many Mixnodes on average she needs to strategically 
    compromise until 
    she can fully observe the complete route of this message, and this metric is called guessing 
    entropy~\cite{rochet2017waterfilling, guess1994entropy}. In particular, this metric can be interpreted 
    as a 
    worst-case adversarial resource endowment
    to guarantee deanonymizing a given single-message target.
%     that captures the
% required number of mixes required to be compromised for a given message to be
% deanonymized.
\end{itemize}

These measures are not only helpful for the mixnet designer or operator but also meaningful for users wishing to know ``How secure am I if I use the system?''

\subsection{Reference Algorithms}
\label{subsec:reference_algos}
We will empirically evaluate our Mixnet construction algorithm on a
statistically significant number of generated topologies, and compare \sys to three
\textit{reference} construction methods: \textit{BwRand}, \textit{RandRand}, \textit{RandBP}, described below. 
% We evaluate the best adversarial strategy against each of the
% construction methods and compare them under their respectively strongest
% adversary. Security notions in this section are defined assuming a fixed point
% (snapshot) in time, i.e., these security notions are \emph{static}. The next
% section considers temporal dynamic aspects.
\begin{itemize}
	\item \textit{BwRand}: 
		the CS samples Mixnodes from the active pool with the probability proportional to their bandwidth and 
places these Mixnodes into random layers with uniform probability.  This is
		a good proxy for the Nym Mixnet design. Indeed, Nym expects to
		sample nodes based on their stake value, which is expected to
		correlate with bandwidth in their reward system (i.e., staking
		to nodes proportional to their true bandwidth maximizes the
		profit).
    \item \textit{RandRand}: CS samples 
the active pool uniformly at random and places the Mixnodes into a layer uniformly at random. 
    \item \textit{RandBP}: CS samples the active 
pool uniformly at random and assigns Mixnodes into each layer with the Bin-packing placement algorithm.
\end{itemize}

%Here the random placement is implemented by a simple approach ---placing mix $i$
%according to the value of $L_i \leftarrow R_i \bmod{3}$ in which $R_i$ is
%the random number generated in round $i$. 

% \xinshu{Add footnote, We abstract Nym as BwRand but not intention to compare with NYm.}

% \subsection{Client Enumeration and Mixnode Adversary}
% \xinshu{the rationale of considering this type of adversary}

\subsection{Adversary Modeling}\label{subsubsec:security_methodology} 
In our simulation, 
% We empirically 
% evaluated the security of Bow-Tie through a series of analyses on generated
% Mixnets. 
we consider one adversary who wants to 
deanonymize messages by optimizing the use of Mixnodes and bandwidth resources. We model 
the adversary bounded by two elements: the number of Mixnodes 
available to the adversary $m$, and the bandwidth available for each node $b^m_
i$, $i \in [1, m]$. An adversary is assumed to control a certain fraction $\alpha$
of the total bandwidth resources, with a resource budget $B_m$ 
such that any combination of $m$ and $b^m_i$ that meets the constraint $B_m \geq \sum
_{i=1}^{m} b^m_i$ can be applied. In our simulation, a candidate mix pool consists of 
$1000$ benign nodes and $m$ malicious nodes. The total bandwidth of candidate pool is $P_{bw}\,\approx\,$$11400$\,MBps including $2280$ MBps malicious bandwidth ($\alpha = 20\%$) such that honest Mixnodes are the majority. 
% Let malicious fraction $\alpha = 0.2$, .

\subsection{Adversary Resource Allocation}
\label{subsec:best_strat}
% \xinshu{4) It would be nice to see actual results for the suggested placement strategies. Don't know how.}

\begin{figure*}[th]
	\centering
	\begin{subfigure}{.5\columnwidth}
		\centering
		\includegraphics[width=0.95\textwidth]{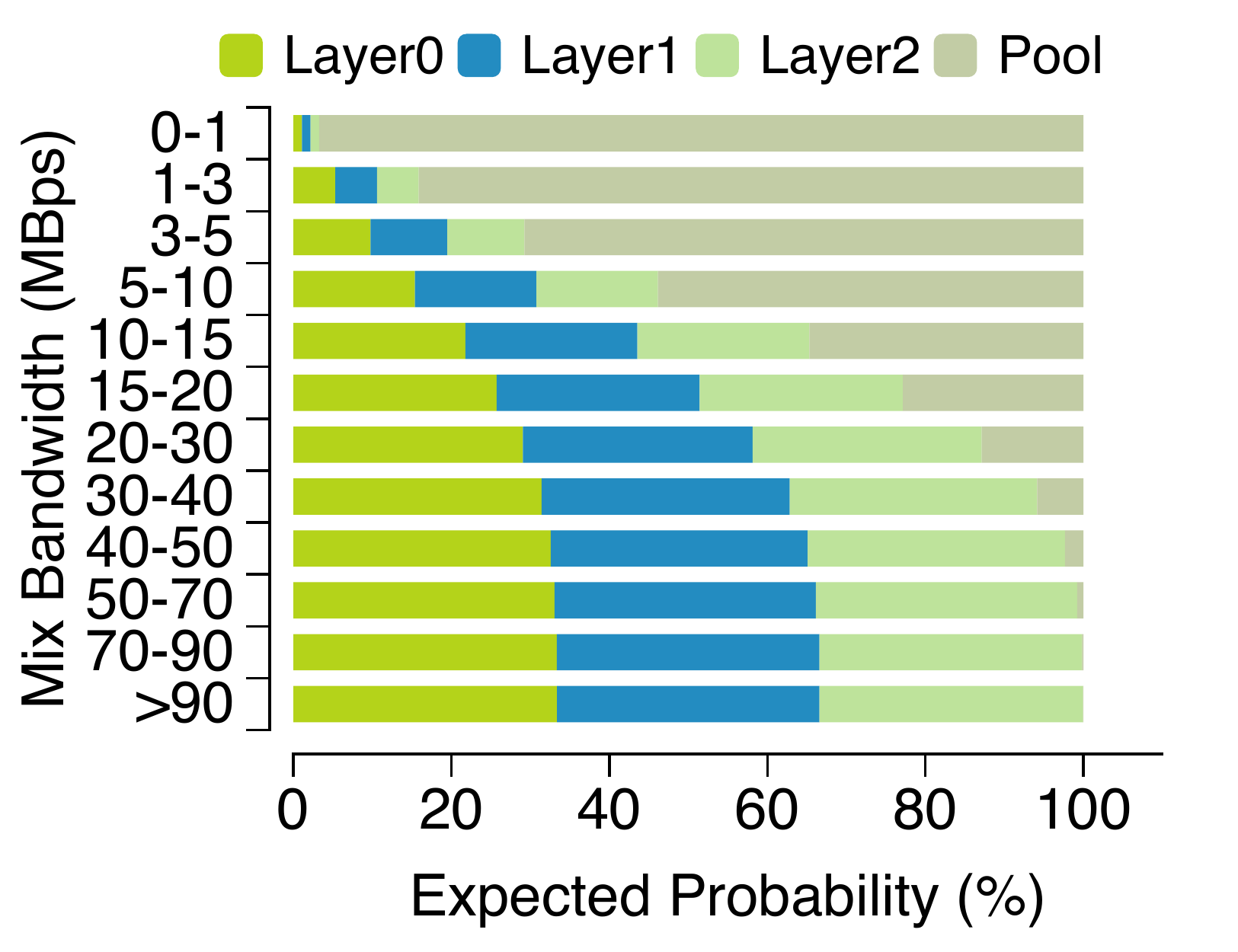}
		\caption{ BwRand}
		\label{subfig:shape_bwrand}
	\end{subfigure}%
	\begin{subfigure}{.5\columnwidth}
		\centering
		\includegraphics[width=0.95\textwidth]{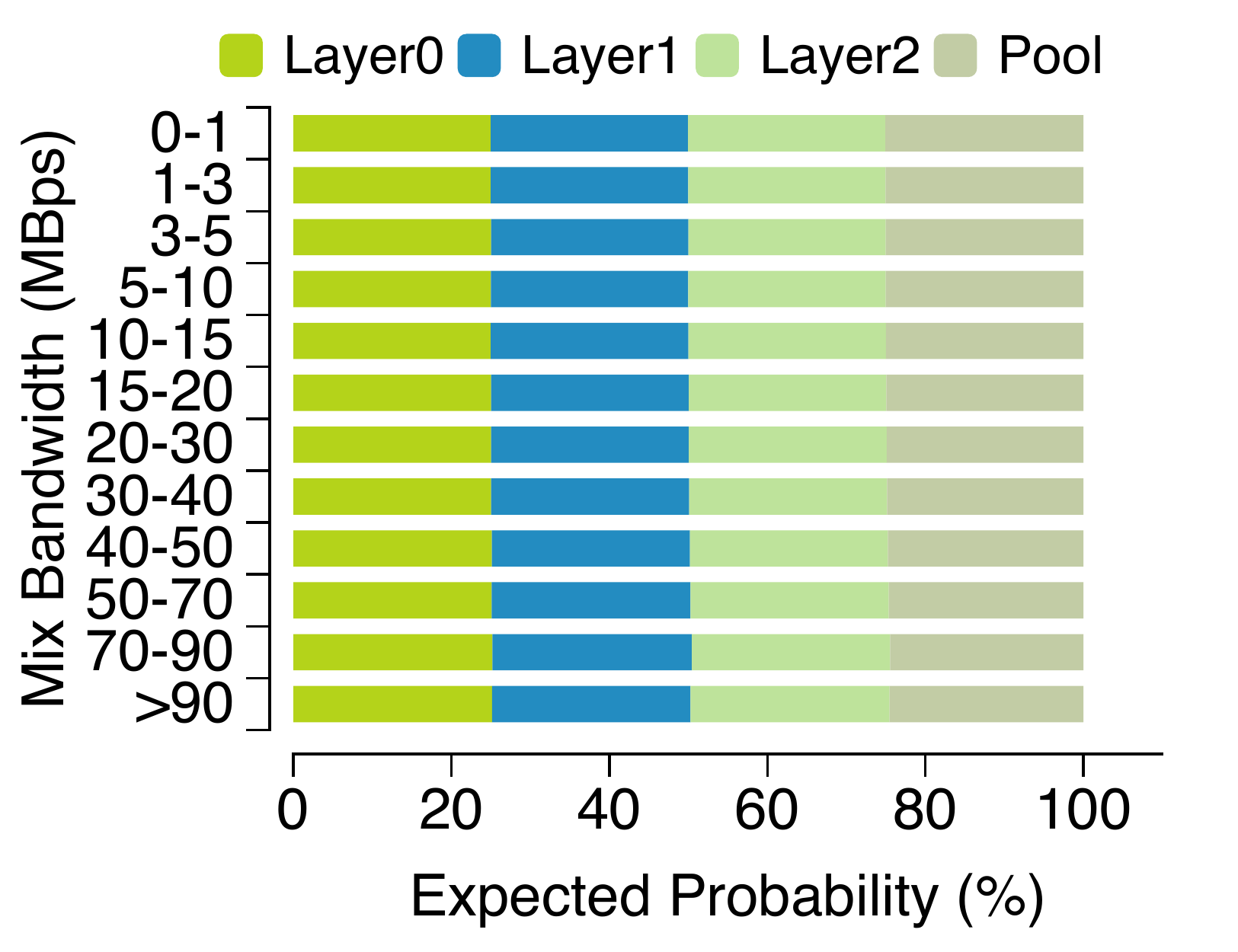}
		\caption{ RandRand}
		\label{subfig:shape_rr}
	\end{subfigure}
	\begin{subfigure}{.5\columnwidth}
		\centering
		\includegraphics[width=0.95\textwidth]{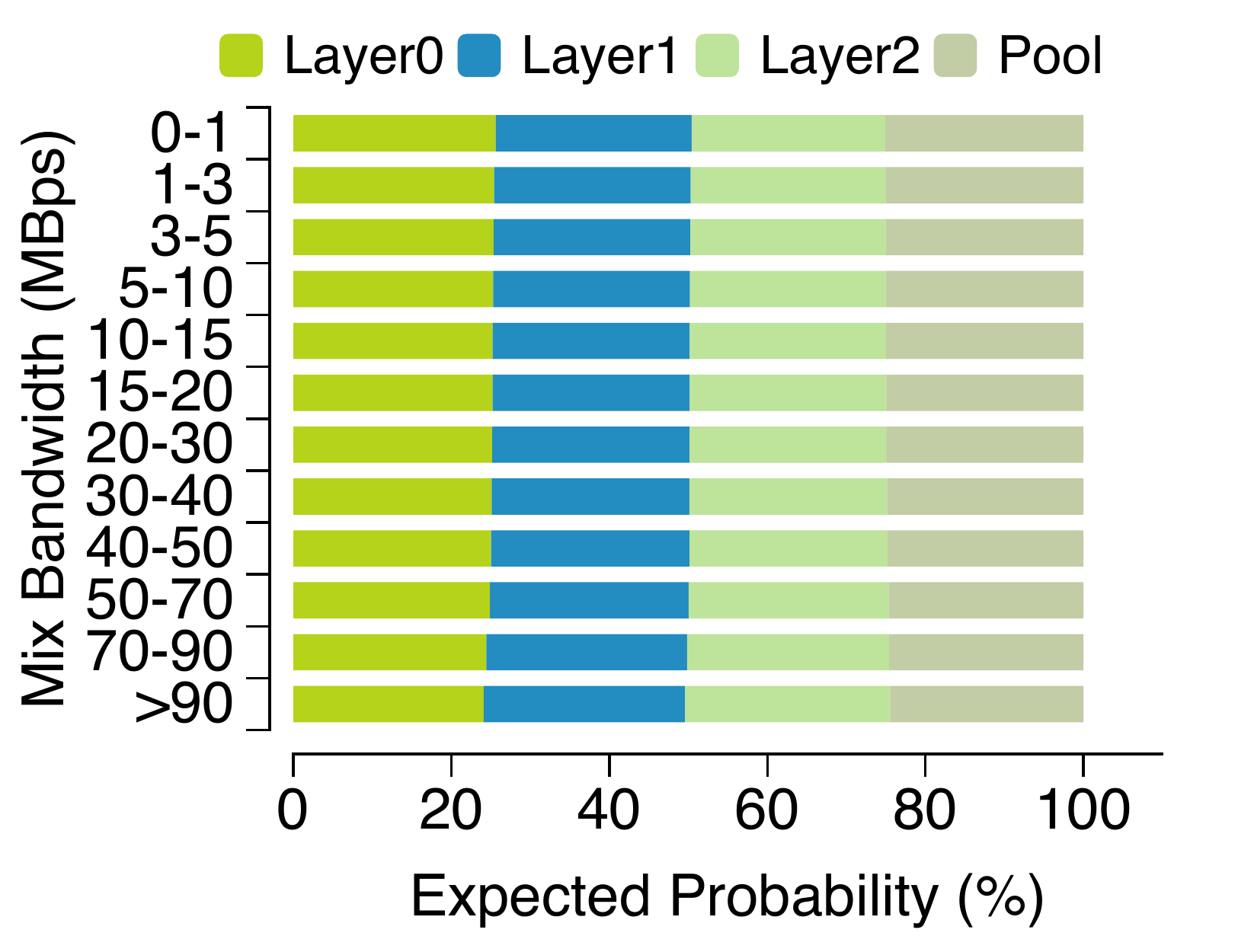}
		\caption{ RandBP}
		\label{subfig:shape_rlp}
	\end{subfigure}
	\begin{subfigure}{.5\columnwidth}
		\centering
		\includegraphics[width=0.95\textwidth]{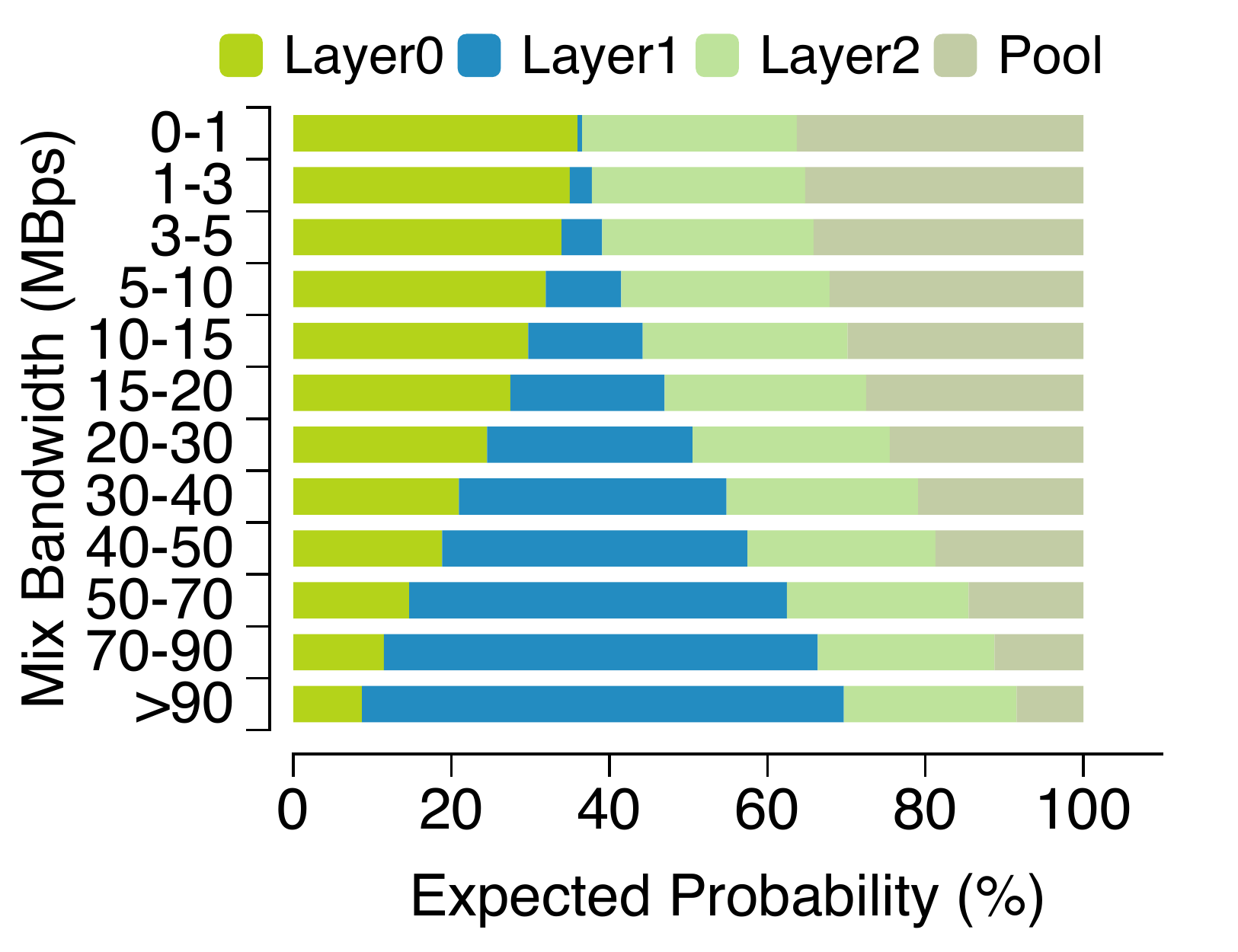}
		\caption{ Bow-Tie}
		\label{subfig:shape_bowtie}
	\end{subfigure}
  \caption{Expected probability to fall in Layer $i$ depending on the Mix
    capacity, with sampling threshold $h=0.75$. Pool is the expected
    probability to stay within the candidate pool and not participate.}
	\label{fig:layer_shape}
\end{figure*}

The adversary must determine how best to allocate his bandwidth to maximize the
compromised fraction of paths. Since the same Mixnode cannot be chosen twice,
he must run at least 3 Mixnodes for a 3-layer Mixnet.  The crucial insight here
is that the adversary has knowledge of what algorithm is being run to establish
topologies and can distribute its total budget  as a particular number of nodes
and bandwidth that would maximize its chances. That is, having a similar
presence in each layer would maximize deanonymization, and the adversary needs
to determine its resource endowment to achieve it.

To answer this adversarial question and thoroughly model the adversary, we ran
numerous experiments using the \texttt{mixnets topology generator}
(Section~\ref{sec:MTG}) to empirically investigate how topological construction
algorithms shape the network. We statistically derive, over $200$\,K runs with
$h = 0.75$, the probability of Mixnodes' placement into different layers. Each
Mixnode will either go to one layer of the Mixnet or remain in the candidate
pool.

The results are displayed in Figure~\ref{fig:layer_shape}, from which we can
infer appropriate allocation strategies for the adversary. We can see
(Figure~\ref{subfig:shape_bwrand}) that BwRand's clear preference for bandwidth
is in favour of big Mixnodes (especially with bandwidth no less than
$70$\,MBps), which will be assigned into three layers evenly.
Figure~\ref{subfig:shape_rlp} show that there is a $25\%$ chance that
each Mixnode will be placed into one of four positions with RandRand or RandBP. Bow-Tie (Figure~\ref{subfig:shape_bowtie}) shows a different position distribution, where Mixnodes with bandwidth in between $20-30$ MBps has the same chance of being placed into any layers. Thus allocating malicious bandwidth resources evenly across a number of nodes makes sense for the adversary.

When generating Mixnet topologies according to different construction designs,
we consider an adversary who has individual resource allocation strategy
respectively. An adversary will generate the following malicious Mixnodes:
nodes with $71.25$ MBps against BwRand, nodes with $11.75$ MBps against RandRand
and RandBP, and nodes with $20.72$ MBps against Bow-Tie. Note that we also test the
compromised fraction of paths with varied capacity (ranging from $1$ MBps to
$150$ MBps\footnote{Allocating too much bandwidth to one node is not realistic
due to the CPU cost of the public-key encryption within each Mixnet packet, so
we set the upper bound of malicious mix to $150$ MBps.}) of equal-size
malicious nodes and the results supports our choices of best allocation
strategy (please refer to the  Appendix~\ref{app:resource allocation}).

%% file: 5_behaviour.tex
\section{Empirical Analysis of Bow-Tie}
\label{sec:5_behavioural}
We now evaluate the security of the Mixnets with respect to the metrics and adversaries in 
Section~\ref{sec:evaluation methodology}. 
To do so, we develop two tools: \texttt{mixnets topology generator (MTG)} to produce the reference and 
\sys topologies and \texttt{routesim} to evaluate the topologies on their expected security metrics of 
typical dynamic email-like usage. We conclude by investigating the necessity of Bow-Tie's guard layer 
and client-side guard-logic. 

\subsection{Tools}
\label{subsec:routesim}

\subsubsection{MTG}
\label{sec:MTG}
We implemented a scalable Mixnet topology generator incorporating the four mixnet construction 
algorithms in Python. We use Gurobi 
optimizer \cite{gurobi2018gurobi} to solve the linear bin-packing 
optimization problem~\cite{scheithauer2017introduction}.
% Our source code is 
% avaiable online\footnote{\url{https://anonymous.4open.science/r/MixnetConstructionSimulator-FD0F/}}.
The bandwidths of Mixnodes are generated by fitting to the bandwidth distribution of 
Tor relays from its historical data~\cite{CollecTor}. We use an R package~\cite{delignette2015fitdistrplus} 
to fit the 
bandwidth data captured from Tor consensus documents and server descriptors 
from January $2021$ to March $2021$.
Among three common right-skewed distributions~\cite{cullen1999probabilistic} 
we choose the \textit{gamma distribution} as the best-fitted via maximum likelihood estimation (MLE) method.

\subsubsection{Routesim}
% In realistic Mixnets usage, users are expected to send more than one message, therefore
% evaluating anonymity only based on entropic definitions leaves several
% dimensions to the problem unexplored. For this reason, we need a new tool and additional
% metrics to capture these other dimensions to user anonymity.

To enable our evaluation of time to first compromise metric based on realistic Mixnets usage, we 
implement 
\texttt{routesim} to support the dynamic, multi-message 
user scenario, aiming at estimating user's resilience against client enumeration.
\texttt{routesim} applies a Monte Carlo method to sample a user's usage distribution and simulate the 
user's expected anonymity impact. For each sample simulation, it takes the message timings and sizes 
following a communication pattern provided by the user, the Mixnet's topology generated by \texttt{MTG} 
for each epoch, and two families of mixing protocol interactions (recipient-anonymous and non 
recipient-anonymous, described in Section~\ref{subsec:proto-inf}) as the input, and outputs the trace of 
all messages that are produced and transmitted through the network.
\texttt{routesim} is written in \emph{Rust}, scales with the number logical
processors, has a low-memory footprint and is designed to be easily extensible
for new client models and probabilistic events to capture.  It can simulate
statistically relevant durations (e.g. months) of a given client behavior in a few minutes on a
regular laptop, i.e. it is usable on low budgets.

\subsection{Analysis}
\subsubsection{Time to First Compromise}
%For any given selected topology, the user's interactions
%with the mix network play a determinant impact on their anonymity, especially
%for mixing strategies (such as continuous-time) that \emph{leak over time.} We consider the expected 
%security of each user over a period of time. 

%We use the metric \emph{time to first
%compromise} route, 
We assume a simple client that sends one
message through the Mixnet every $5$ to $15$ minutes at random within this
interval and we model $10,000$ such
clients. We use \texttt{routesim} to conduct simulations and obtain the distribution of time to first 
comprimsed message. 
%from Monte Carol method. 
The network churn rate between each epoch is 3\% for each simulation. The epoch value is
set to 1 hour; i.e., at each epoch, the network topology is refreshed according
to the topology sampling and placement algorithms introduced in
Section~\ref{sec:4_topologies}. The choice of one hour copies Tor's consensus
document renewal.
%\xinshu{There are two kinds of epoch length: 1h-epoch for some expriments and 1day-epoch for loopix and rdv comparison. What about email experiment? Reviewer asks about why this parameter was chosen this way?}

% \texttt{routesim} can simulate activities from both families and eventually
% output a distribution function of a probabilistic event. Such an event can be
% \textit{Time to First Compromised Message} and \textit{Number of Messages Sent
%   Until Compromised} and can be considered anonymity metrics. While the second
% metric can be obtained analytically by modelling route compromise as Bernoulli
% trials, and applying Chernoff's bound~\cite{wright2004predecessor}, the first metric can
% only be obtained from the Monte Carlo method. The Monte
% Carlo method in \texttt{routesim} also gives the second metric for free.

\begin{figure}[!h] 
	\centering
	\begin{subfigure}[b]{0.225\textwidth}
		\centering
		\includegraphics[width=1.05\columnwidth]{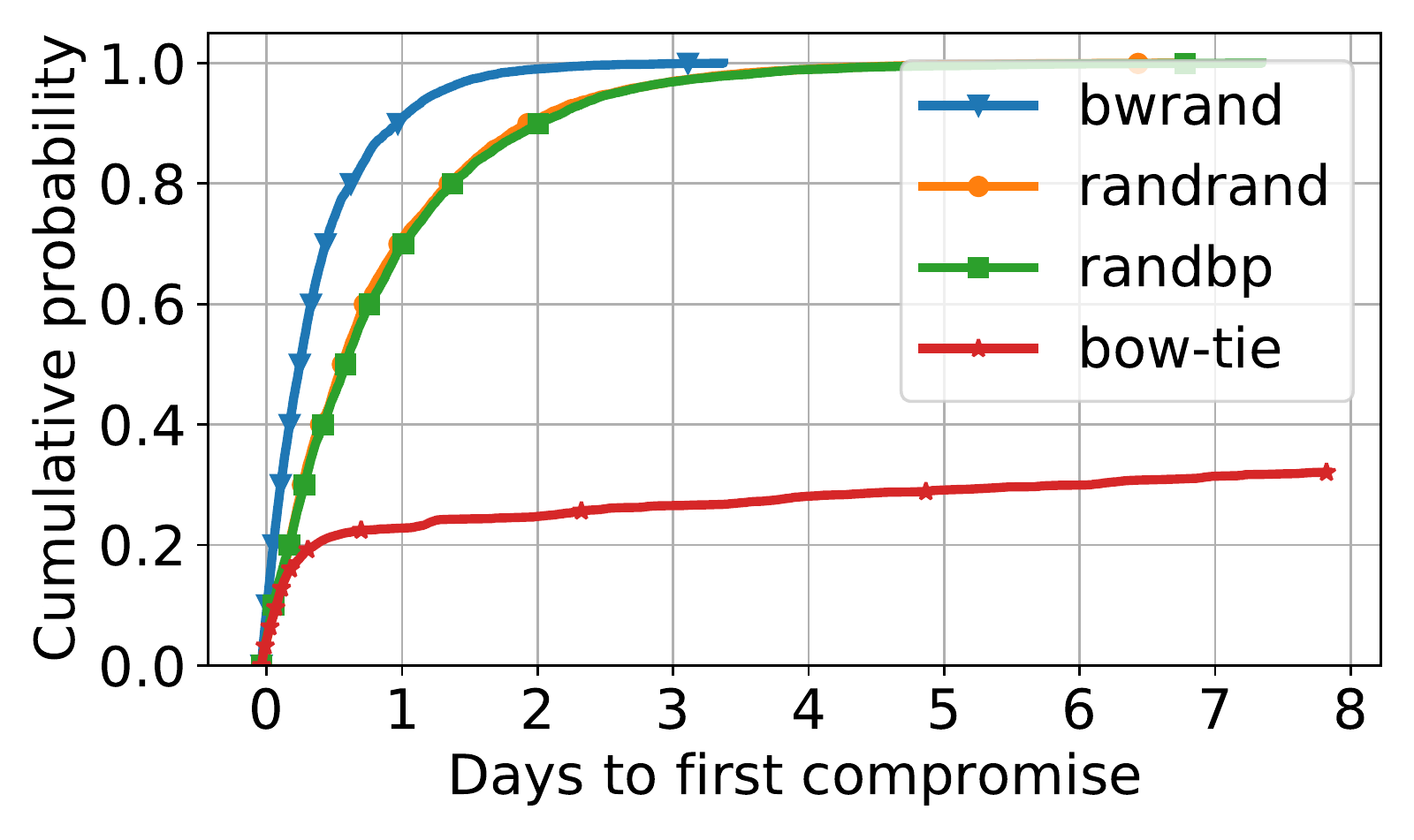}
    \caption{Time to first compromised message.}
		\label{subfig:full_dynamic_time}
	\end{subfigure}
	\hfill
	\begin{subfigure}[b]{0.225\textwidth}
		\centering
		\includegraphics[width=1.05\columnwidth]{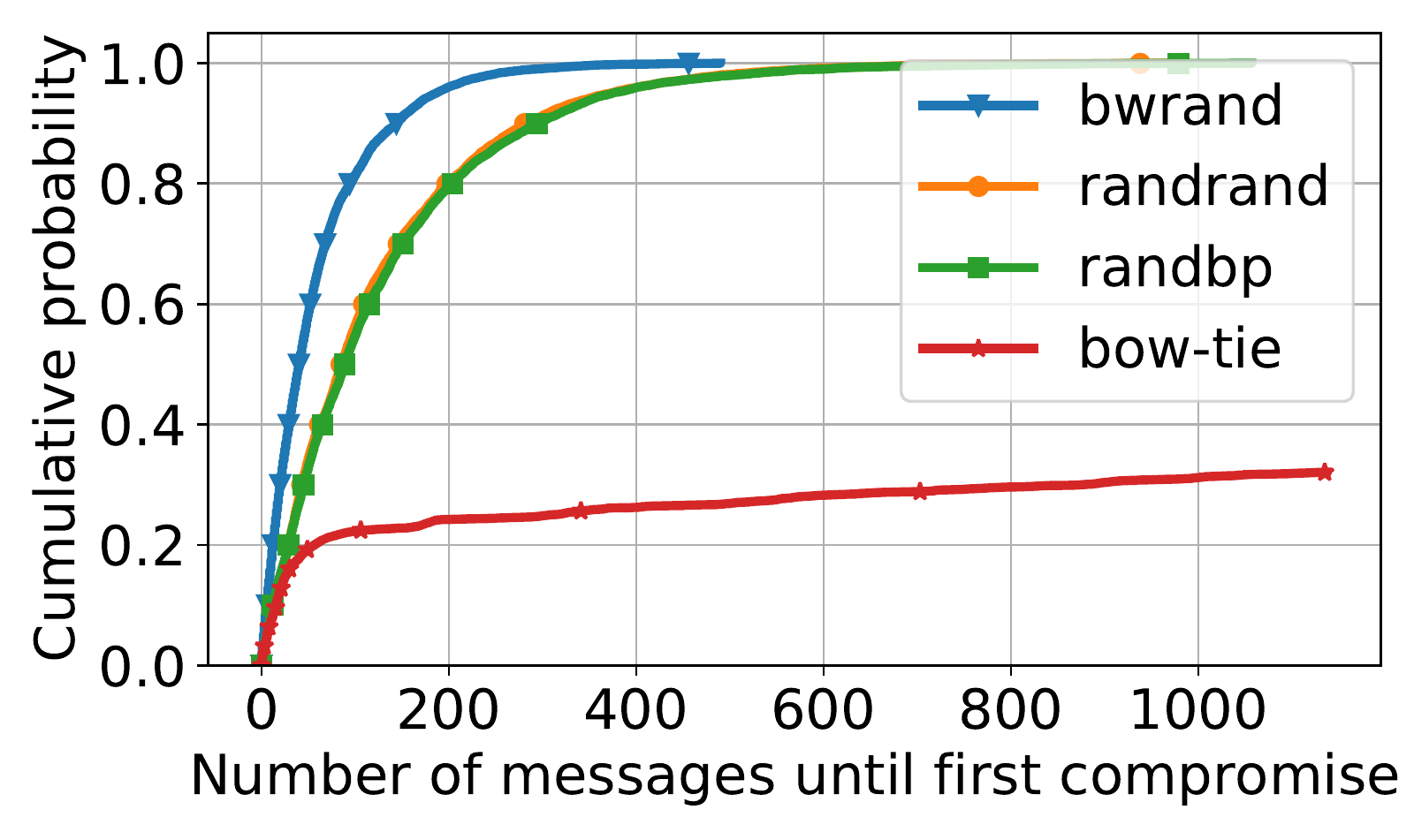}
    \caption{Number of messages until first compromised message.}
		\label{subfig:full_dynamic_count}
	\end{subfigure}
  \caption{Empirical distribution of how much time/how many messages before a user's message traverses over a fully compromised path since first usage. We model a user sending one message every 5 to 15 minutes at
  random.}
	\label{fig:full_dynamic}
\end{figure}

\textbf{Results.} Figure~\ref{fig:full_dynamic} shows the CDF
for the event that a user's message first traverses over a fully compromised path.
% samples (i.e., 10,000 clients generating
%  messages probabilistically through the pattern following the ``simple'' distribution). 
For the reference designs, the client is expected to use a fully compromised 
route extremely fast since each message has the potential to go over \textit{any} 
%(bandwith-weighted for all topologies) from the set of 
of the potential routes and the users will expose themselves to many Mixnodes, including adversarial 
ones. 
We can see (Figure~\ref{subfig:full_dynamic_time}) that with all three 
reference designs there is more than $80\%$ chance of 
deanonymization of at least one message within 2 days by an adversarial Mixnode and the median time to 
full compromise is less than 0.7 days.
%We also see that this risk arises steadily over time.
By looking at the distribution of messages sent ( Figure~\ref{subfig:full_dynamic_count}) for the 
reference designs, the median number of messages sent (for the ``simple'' client model) before 
compromise is 100. In contrast, Bow-Tie enjoys a significantly longer time and higher number of 
messages sent until first compromise.

\subsubsection{Compromised Fraction of Paths}
\label{subsubsec:fully compromised paths}
We now evaluate how many network paths the adversary may control by considering the fraction of 
compromised paths metric. 
%That an adversary compromises some paths is significant, but how many he compromised is just as 
%important. We use the metric of compromised fraction of paths to provide the information of the 
%network 
%itself. 
Recall that a path or route within Mixnets is compromised if the entire route is composed by malicious 
Mixnodes. Thus, we set the compromised fraction of paths $F_b$ in a 
stratified $l$-layer Mixnet using bandwidth-weighted message forwarding as

\begin{equation}
\label{equ:fully comp rate bw}
F_b = \prod_{i=1}^{l} \frac{\text{Amount of Malicious Bandwidth in Layer $i$}}{\text{Amount of Bandwidth 
in 
Layer $i$}}.
\end{equation}

\begin{figure}[!h]
	\centering
	\begin{subfigure}{0.225\textwidth}
		\centering
		\includegraphics[width=\textwidth]{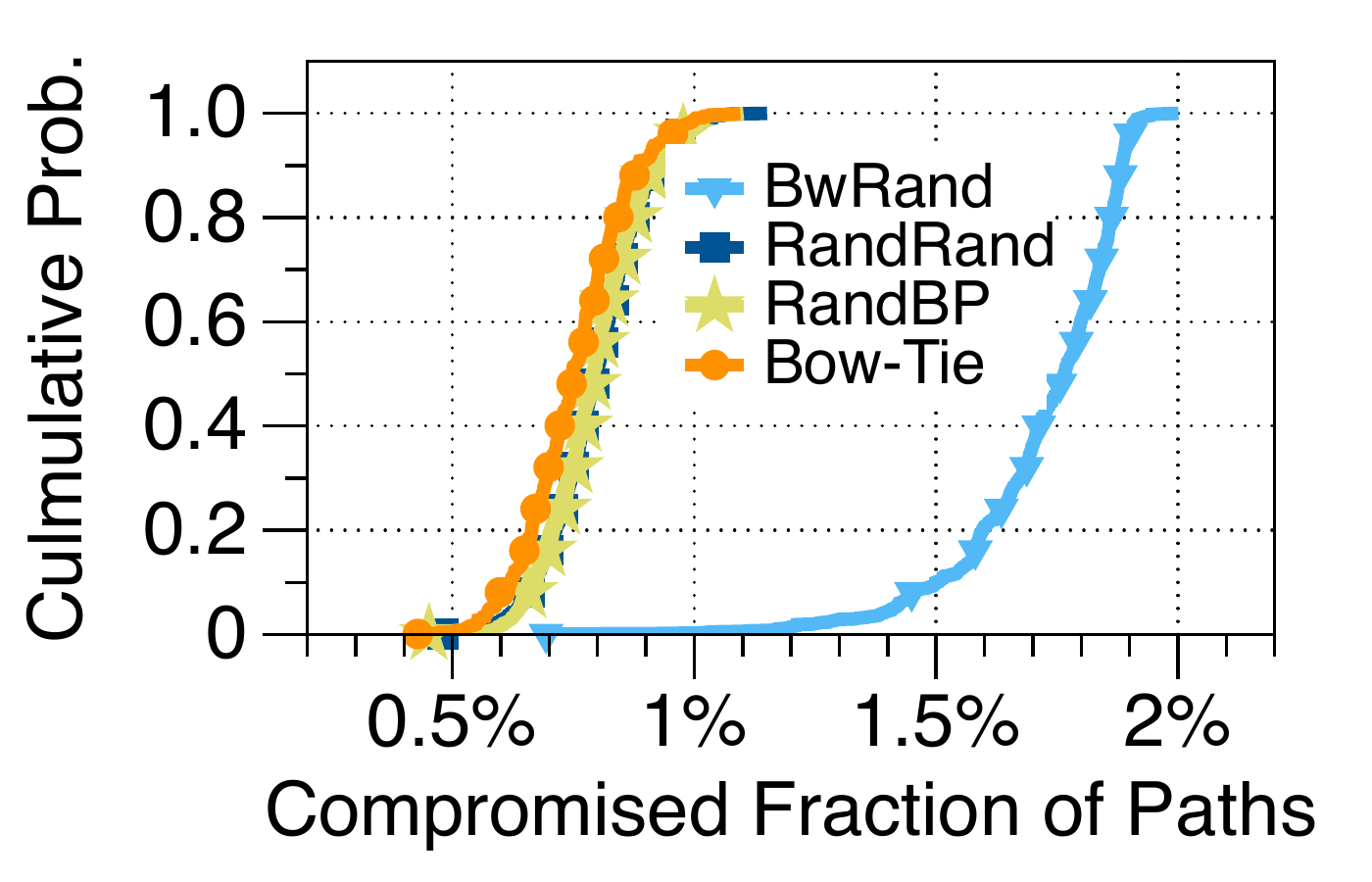}
		\caption{Probability distribution on compromised fraction of paths with $h=0.75$.}
		\label{subfig:best_bw_075}
	\end{subfigure}
	\begin{subfigure}{0.225\textwidth}
		\centering
		\includegraphics[width=\textwidth]{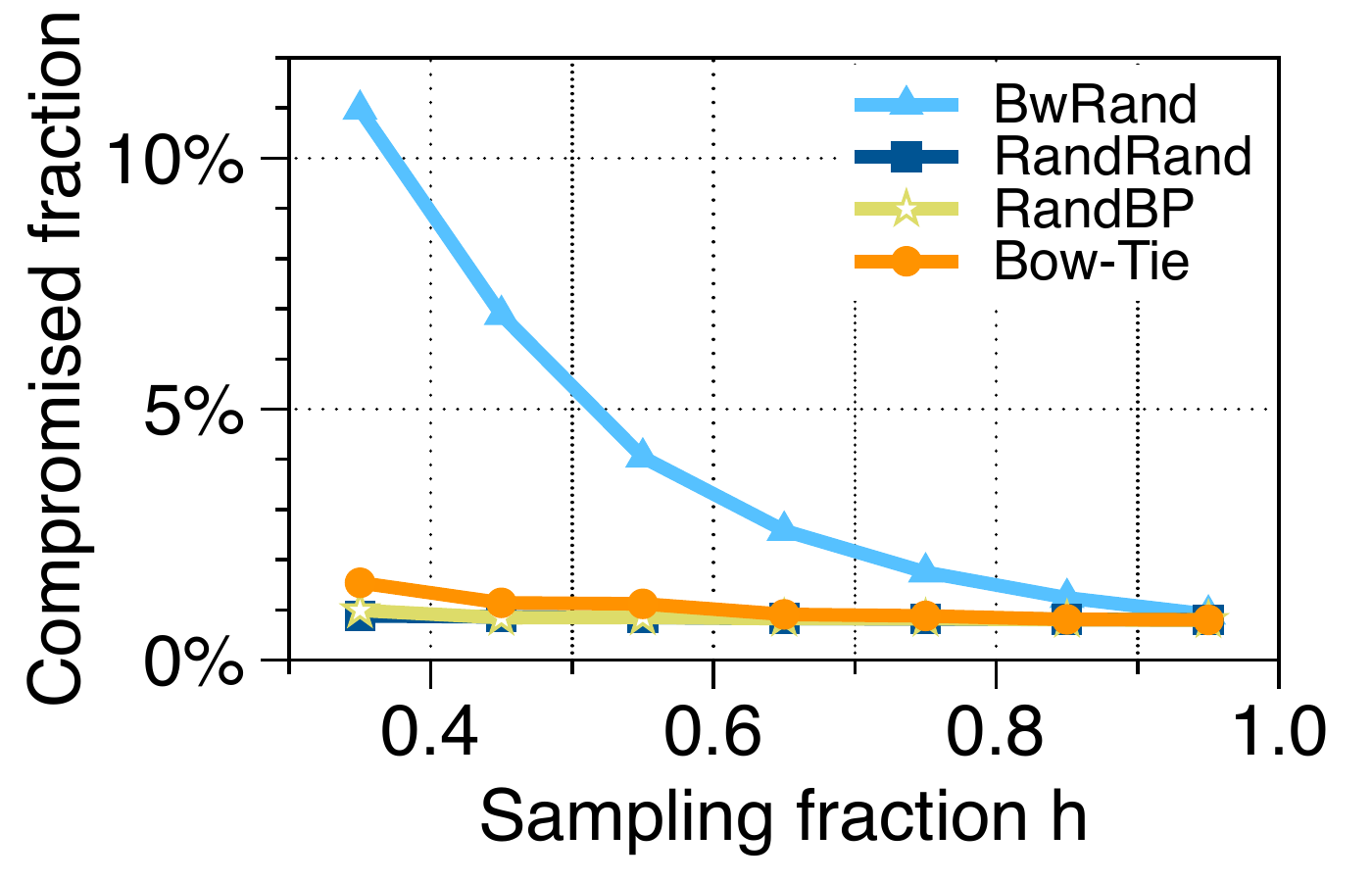}
		\caption{Average compromised fraction of paths for $h$ between $0.35$ to $0.95$.}
		\label{subfig:bw_all_a02}
	\end{subfigure}
  \caption{Empirical distribution on compromised fraction of paths and empirical average compromised fraction. 
%   We model the sampling fraction $h$ from $0.35$ to $0.95$, i.e., a subset nodes, with bandwidth fraction---$h$ of the total bandwidth, are selected to operate in the Mixnet. The adversary controls $20\%$ bandwidth resources of the total.
  }
%   average compromise rates and their CDFs for bandwidth-weighted
% 	path selection (a-d), and random path selection (e-h), when the intelligent adversary
%   uses the best resource allocation policy explored in
%   Section~\ref{subsec:best_strat}.}
	\label{fig:compromised path}
\end{figure}

% \xinshu{fix xlable of figure7a, and ylable of figure7b to compromised fraction of paths. Thick line, and smaller marks (stars, triangles)}

We empirically evaluate this metric, statistically derived over $1000$ runs, with simulations that construct 
Mixnets using Bow-Tie and reference algorithms. Adversarial bandwidth is set to $20\%$ of the total 
network bandwidth. 

\textbf{Results.} 
%Figure~\ref{fig:compromised path} shows how many paths an adversary may fully compromise. 
Figure~\ref{subfig:best_bw_075} shows that, when $h=0.75$, there is more than a $99\%$ chance 
of compromising less than $1\%$ paths using Bow-Tie, and the Rand- reference algorithms.
% by an adversary with the 
%nodes that count for $75\%$ of the total bandwidth resources operating in the 
%Mixnet, 
In contrast, in BwRand the adversary can compromise upto $2\%$ paths. This is because selecting all 
nodes by bandwidth in BwRand gives the adversary that intelligently allocates bandwidth an advantage. 
This is not so effective against Bow-tie since the non-guard layers use random placement. 

Figure~\ref{subfig:bw_all_a02} shows the worst-case expected compromise rates, where all malicious 
relays are selected for use under all the values of $h$ considered.  We see that Bow-Tie, RandRand, and 
RandBP have generally low compromise rates across all sampling 
fractions $h$, with Bow-Tie slightly higher (less than $0.05\%$) 
when $h < 0.6$. This is due to the fact that the guard layer is bandwidth weighted, however, the 
non-guard layers minimize an intelligent adversary's optimal allocation strategy. In contrast, as $h$ 
decreases BwRand's compromise rates increase, with $10.9\%$ of paths compromised when $h = 
0.35$. %and $0.97\%$ of paths compromised when $h = 0.95$. 
The compromise rates are generally converging towards a lower value (around $0.08\%$) as $h$ 
increases for all algorithms, which is expected since more honest nodes will enter the active pool and the 
fraction of adversarial relays will decrease.

This raises an interesting question about how to derive and adjust the parameter of sampling fraction 
$h$. The appropriate $h$ should be able to handle all of the incoming traffic without  overloading 
 the majority of Mixnodes, and should limit the number of paths in the network to avoid very thin traffic 
from the perspective of entropy~\cite{guirat1mixnet}. Thus, $h$ should be set as a minimum value that 
satisfies the throughput requirement, based on historical data or reasonable predictions. We leave as 
future work the case when the volume of incoming traffic changes suddenly within an epoch.

\subsubsection{Guessing Entropy}
\label{subsubsec:guessing entropy}
%We now consider an \textit{active external} adversary for the scenario where she is targeting a specific 
%message 
%sent from a particular user. We wonder how many Mixnodes on average she needs to strategically 
%compromise until 
%she can fully observe the complete route of this message, and this metric is called guessing 
%entropy~\cite{rochet2017waterfilling, guess1994entropy}. In particular, this metric can be interpreted 
%as a 
%worst-case adversarial resource endowment
%to guarantee deanonymizing a given single-message target.

% to capture how many Mixnodes that this adversary need to compromise in order to deanoonymize a given message. 
% We consider an active external adversary who is targeting one message sent
% from a specific user and use the metric of guessing entropy 
%  This metric~\cite{rochet2017waterfilling, guess1994entropy} 
% gives the average number of Mixnodes required to be strategically compromised by
% the adversary until she can fully observe the complete route of the target
% message. 
We model the deanonymization of a given message as a guess and let $\mathcal{G}$ represent the total 
number of guesses for
success (i.e. deanonymizing the target message). $\mathbb E(\mathcal{G})$ is computed by
selecting the nodes in descending order of the marginal probability
$p_i$ that the adversary can deanonymize the targeted message when cumulatively
compromising the $i_{th}$ node. Thus, the guessing entropy can be calculated by: 
\begin{equation} 
  \label{equ:guess_entro} 
	\mathbb{E}(\mathcal{G}) = \displaystyle\sum\limits_{i \in |Pool_{active}|} i \cdot p_i, 
% \displaystyle\sum\limits_{i \in I, j \in J}           b_i x_{ij} <= Cy_{j} 
\end{equation}
where $|Pool_{active}|$ represents the number of Mixnodes in the \textit{active
  pool}.

\begin{figure}
    \centering
    	\begin{subfigure}[t]{0.225\textwidth}
		\centering
		\includegraphics[width=\textwidth]{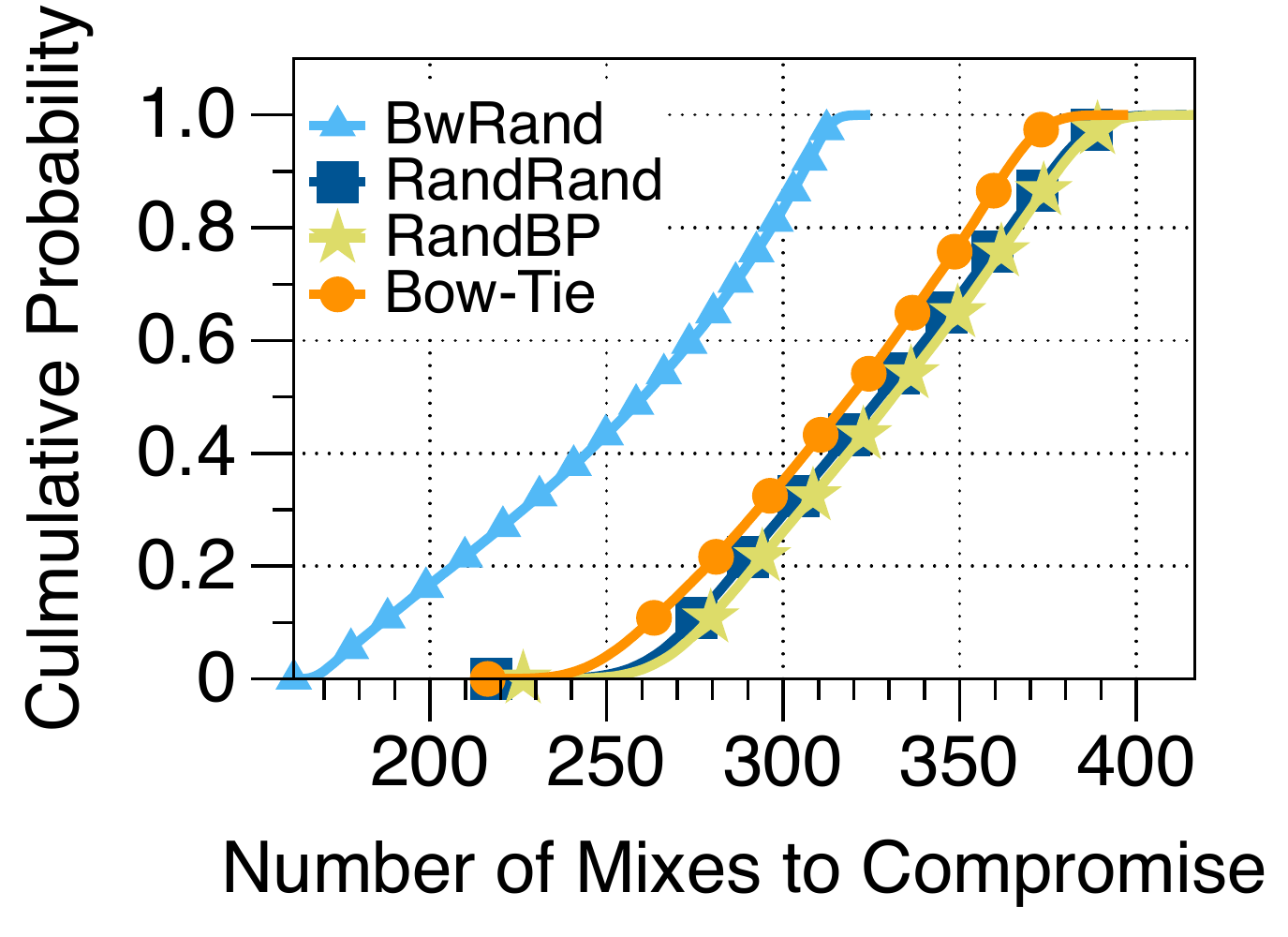}
	\caption{Guessing Entropy}
	% : expected number of Mixnodes the adversary had to compromise to trace a target message.}
	\label{fig:guess_entropy}
	\end{subfigure}
	\begin{subfigure}[t]{0.225\textwidth}
		\centering
\includegraphics[width=\textwidth]{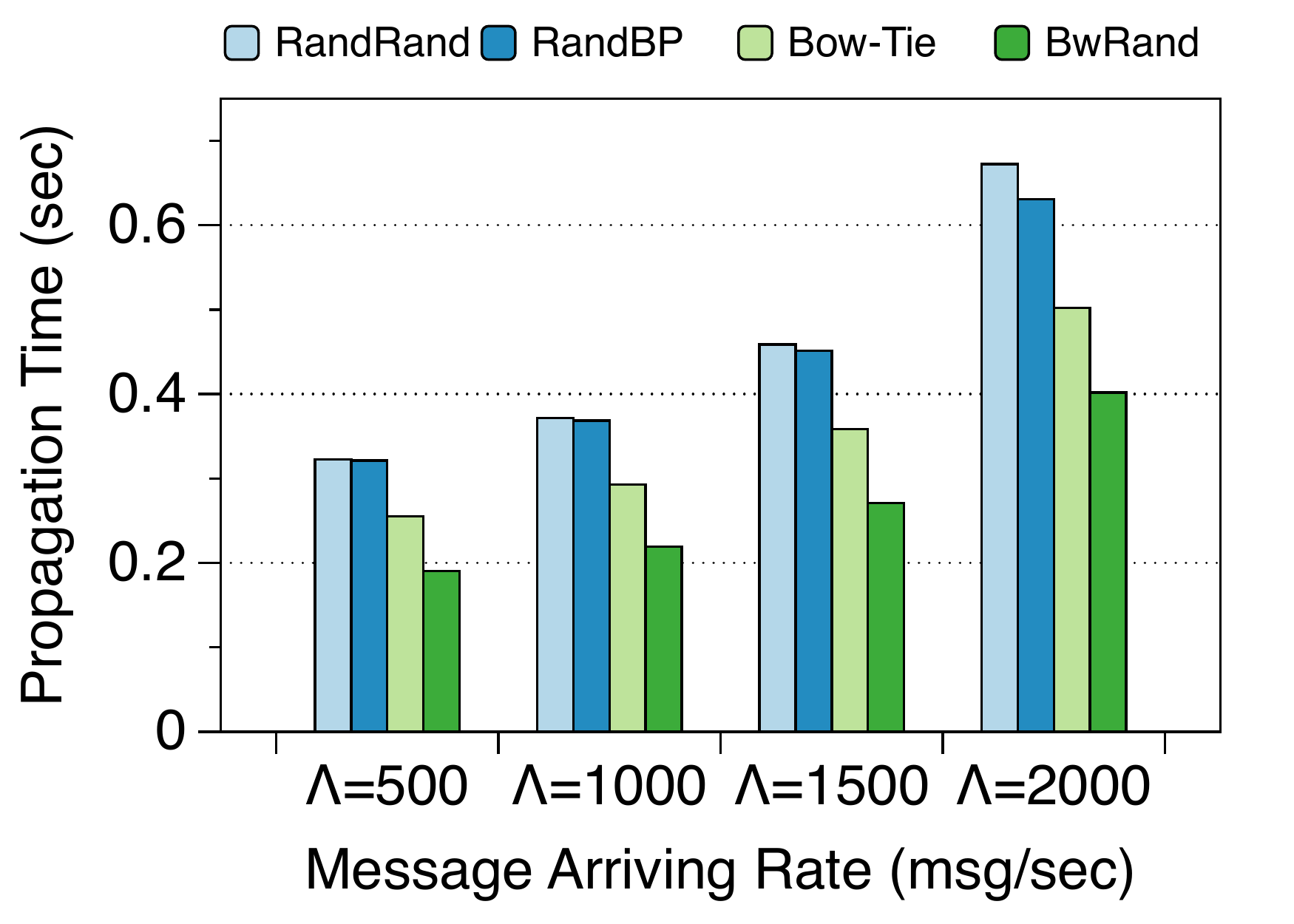}
	\caption{Expected queuing delay}
	% , with message arriving rate $\Lambda$ for the Mixnet.}
	\label{fig:performance}
	\end{subfigure}
    \caption{Guessing Entropy and performance evaluation results. (a) Guessing Entropy: expected number of Mixnodes the adversary had to compromise to trace a target message. (b) Expected queuing delay, with message arriving rate $\Lambda$ for the Mixnet.}
    \label{fig:my_label}
\end{figure}

% \begin{figure}[h!]
% 	\centering
% 	\includegraphics[width=0.27\textwidth]{figures/GE_cdf.pdf}
% 	\caption{Guessing Entropy as a security metric, indicating the average number 
%     of Mixnodes the adversary has to compromise in order to deanonymize a
%     targeted message, in expectation.}
% 	\label{fig:guess_entropy}
% 	% \Description{xx}
% \end{figure}
\textbf{Results.} Figure~\ref{fig:guess_entropy} shows the cumulative guessing entropy value
obtained from $1000$ trials of each topological construction algorithm, for a
network containing $\approx\,1000$ nodes. We can see that the median number of Mixnodes required to 
compromise by an adversary for BwRand is around $250$, while for other three algorithms, the median is 
increased to less than $320$. While \sys is edged out by RandRand, and RandBP, it is significantly more 
secure in the dynamic setting (above) and with better performance, as we shall see next.

% With the BwRand construction
% algorithm, the adversary needs 250 of the most impactful nodes of the network
% on average to deanonymize the single message target in expectation. For the other three
% algorithms, the average number of Mixnodes that the adversary needs to control is
% increased to around $320$.

\subsubsection{Performance Evaluation}
\label{subsubsec:performance}
We measure the \textit{expected queuing delay} (i.e., expected message 
queuing time) based on the topologies generated 
by the \texttt{MTG} with $h = 0.75, \alpha = 0.2$. 
The expected queuing delay is calculated by using a $M/D/1$ queue model~\cite{
murdoch2008metrics}. The input messages of the 
whole mix network can be treated as a Poisson process with rate $\Lambda$. The 
message queuing time for each node is inversely proportional to its
capacity, e.g., for a Mixnode with $b_i$ bandwidth, the average processing 
time for it is $u_i = 1/b_i$. \footnote{We focus on bandwidth-weighted path selection since it performs an 
order of 
magnitude better than random path selection. The interested reader can refer to 
Appendix~\ref{sec:random path selection} for the random path selection performance results.}
% Suppose there are $n$ nodes in one layer, then 
% the expected queuing time in Rd-PS setting for this 
% layer is:

% \begin{equation}
% \label{equ:rand_expect_time}
% T_r = \sum\limits_{i=1}^{n}\frac{n^{-1} u_i 
% (2-n^{-1}u_i\Lambda)}{2(1-n^{-1}u_i\Lambda)}.
% \end{equation}
In bandwidth-weighted path selection, using $U$ to represent the 
total bandwidth of the current layer, the expected queuing time of this layer is:

\begin{equation}
\label{equ:bw_expect_time}
T_b = \sum\limits_{i=1}^{k} \frac{2-\frac{\Lambda}{U}}{2(U-\Lambda)} = k\frac{2-\frac{\Lambda}{U}}{2(U-\Lambda)}.
\end{equation}

% \begin{figure}[h]
% 		\centering
% 		\includegraphics[width=0.26\textwidth]{figures/waittime_bw.pdf}
% 	\caption{Expected queuing delay, with message arriving rate $\Lambda$ for the Mixnet.}
% 	\label{fig:performance}
% \end{figure}

% \textit{Results.}\label{subsubsec:performance_analysis}
\textbf{Results.} Figure~\ref{fig:performance} shows the expected delay due to queuing for a message 
going through the 
Mixnet. Indeed, algorithms that sample using bandwidth
(i.e., BwRand and Bow-Tie) achieve relatively low processing delay and outperform random sampling 
schemes. Compared to BwRand, Bow-Tie sacrifices less than $0.05$ seconds of queuing delay for a 
comparatively higher 
security level (see Figures~\ref{fig:full_dynamic} and~\ref{subfig:best_bw_075}). 

%Another reason that we 
%choose bandwidth-weighted path selection is because 
%random path selection provides much worse performance, where it takes much more time to process 
%even tenth of incoming messages. For more details please refer to Appendix~\ref{sec:random path 
%selection}.

% If the case of Bw-PS, the Mixnet can handle high message
% arrival rates one order of magnitude faster than in Rd-PS.
% These results confirm that Bw-PS contributes to dramatically improved processing speed. 
% using a Bw-PS and Rd-PS, for all
% Mixnet topological construction algorithms presented. 
%It is worth noting that the throughput of Mixnet is bounded by the layer with
%least bandwidth in bandwidth-weighted path selection, and by the layer with
%least number of nodes in random path selection. With the same mix network,

Note that Bow-Tie topologies are also fast to generate: a subsecond cost to both generate the Guard 
layer and to apply the 
bin packing optimization to the other layers. 

\subsubsection{Recap}

Our empirical results in this section confirm that the construction and 
routing of a Mixnet is characterised by a security and performance 
trade-off. Taken together, the results for these 
metrics show that Bow-Tie provides a high level of protection for users' 
anonymity in a dynamic and realistic setting
%and a well-rounded defence against two types of adversaries in general, 
with a relatively small sacrifice in performance.

% Bow-Tie finds 
% a good balance that provides high level 
% security and enables good performance and efficiency at the same time.

% BwRand 
% provides good performance yet it is vulnerable to adversarial 
% manipulation. RandRand and RandBP are essentially equivalent but with 
% different placement implementation methods and differing 
% performance. Both of them can achieve good security levels in 
% Bw-PS (although they are the worst in Rd-PS), yet their performances are relatively bad. Bow-Tie finds 
% a good balance that provides high level 
% security and enables good performance and efficiency at the same time. By 
% comparing all these results, we find that Bw-PS 
% affords better security and load-balance with Bow-Tie (as discussed in Section
% ~\ref{subsubsec:compromised traffic} and Section
% ~\ref{subsubsec:performance_analysis}).

\subsection{The necessity of both client guard logic and guard layers}
\label{subsec:necessity of guard}

%<<<<<<< HEAD
\subsubsection{Turn off Client Guard-logic for Bow-Tie?}
A natural question is does the guard layer by itself (i.e. where the client does
\textit{not} maintain a guard list) provide a high level of protection. To
%=======
%\subsubsection{What if we turn off the client guard logic for Bow-Tie?}
%A natural question is do guard layers by themselves (i.e. where the client does
%\textit{not} maintain a guard list) provide sufficient level of protection. To
%>>>>>>> 1cfb266f14d1aa56602a48a7c97719a08b4310c8
answer this question we turn off the clients' guard list maintenance logic
while they use the network, but keep Bow-Tie's other aspects the same (i.e. \sys
still produces a guard layer).

\begin{figure}[!h] 
	\centering
	\begin{subfigure}[b]{0.225\textwidth}
		\centering
		\includegraphics[width=\textwidth]{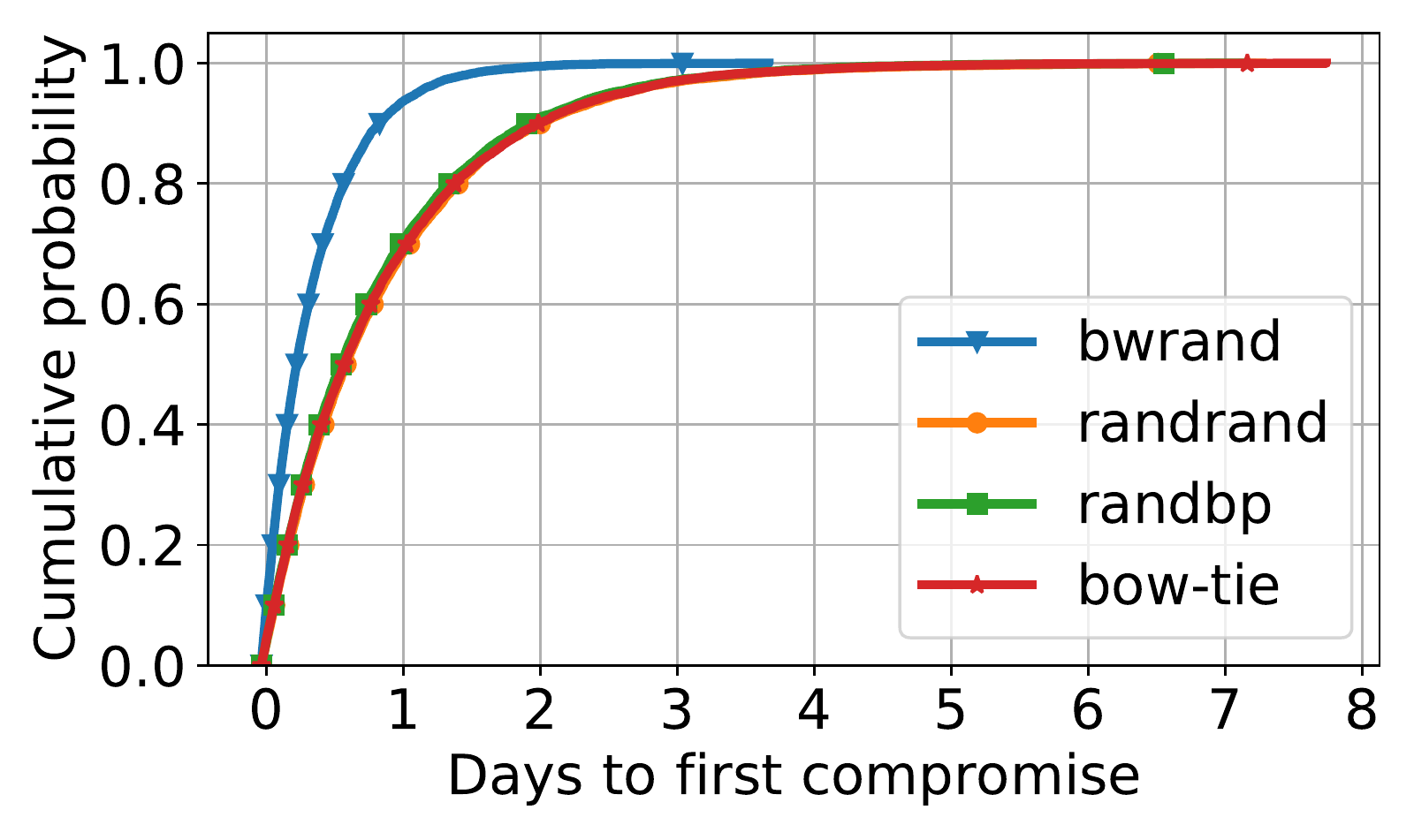}
    \caption{Time to first compromised message.}
		\label{fig:noguard_time}
	\end{subfigure}
	\hfill
	\begin{subfigure}[b]{0.225\textwidth}
		\centering
		\includegraphics[width=\textwidth]{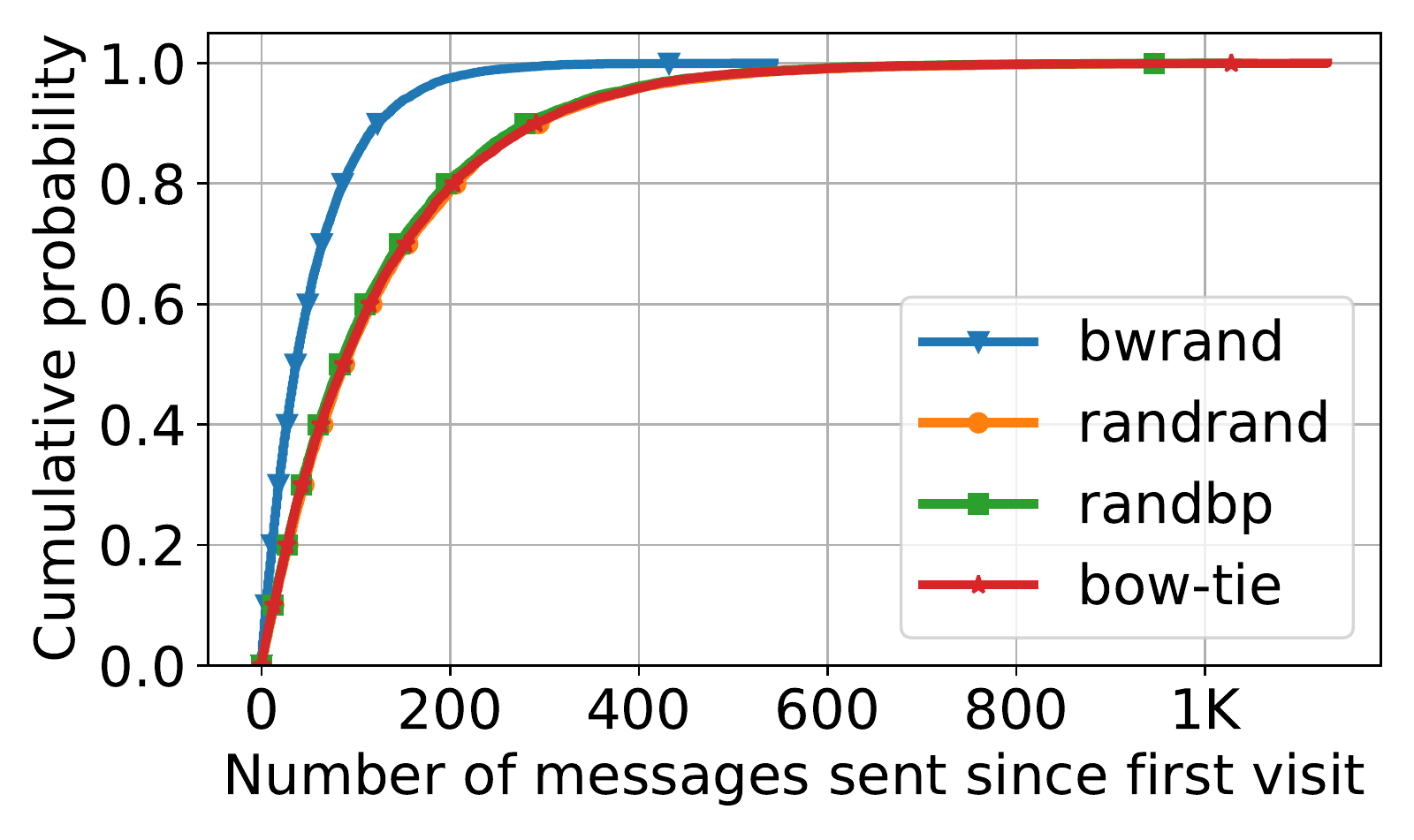}
    \caption{Number of messages until first compromised message.}
		\label{fig:noguard_count}
	\end{subfigure}
  \caption{Comparison between reference algorithms and Bow-Tie without client side guard logic.}
	\label{fig:dynamic_noguardlogic}
\end{figure}

As we observe in Figure~\ref{fig:dynamic_noguardlogic}, a guard layer by 
itself has reduced security at a comparable level to those schemes without guard layers (i.e. RandRand,
RandBP, and BwRand), although Bow-Tie is still slightly better. 
Nevertheless, 
%no matter how well engineered the fundamental topologies, 
the client is expected to use a fully 
compromised route extremely fast since each message is sent at random (bandwidth-weighted here), 
and the users will expose themselves to many
Mixnodes. This implies that users should not explore all potential routes, which is the exact effect of the 
client guard-logic.

% static results
% \begin{figure}[!h] 
% 	\centering
% 	\begin{subfigure}[b]{0.225\textwidth}
% 		\centering
% 		\includegraphics[width=1.05\columnwidth]{figures/ttfb_static_simple.pdf}
%     \caption{Time to First Compromise Event}
% 		\label{fig:static_topology}
% 	\end{subfigure}
% 	\hfill
% 	\begin{subfigure}[b]{0.225\textwidth}
% 		\centering
% 		\includegraphics[width=1.05\columnwidth]{figures/counts_messages_simple_topologies_noguard.pdf}
%     \caption{How many messages sent before one traverses over a fully compromised path}
% 		\label{fig:static_topology_count}
% 	\end{subfigure}
%   \caption{We model a user sending one message every 5 to 15 minutes at
%     random, and record the first occurrence of a fully compromised path.
%     The client side Guard logic is disabled for this experiment.}
% 	\label{fig:static}
% \end{figure}

\begin{figure}[!h] 
	\centering
	\begin{subfigure}[b]{0.225\textwidth}
		\centering
		\includegraphics[width=1.05\columnwidth]{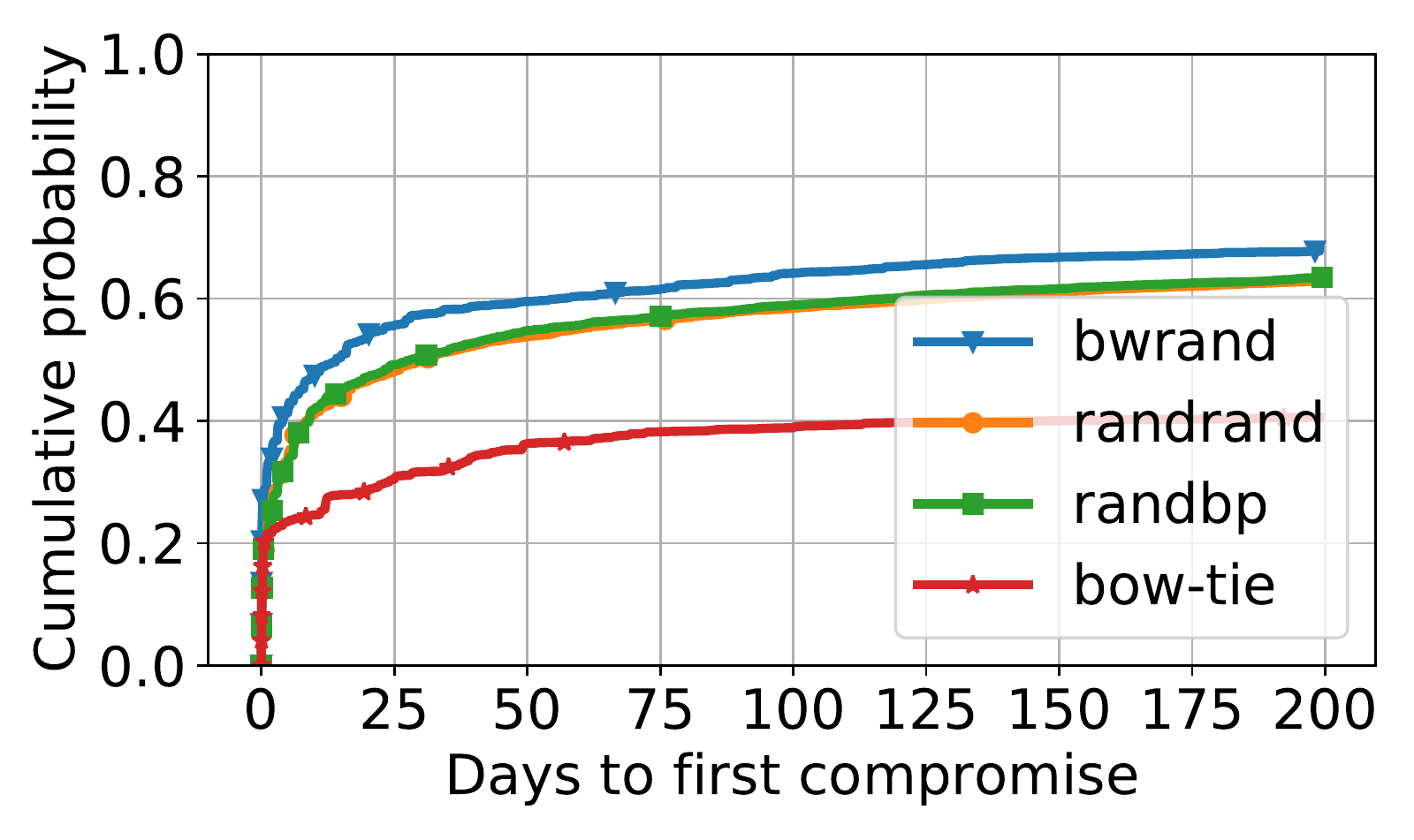}
    \caption{Time to first compromised messag.}
		\label{fig:withguard_time}
	\end{subfigure}
	\hfill
	\begin{subfigure}[b]{0.225\textwidth}
		\centering
		\includegraphics[width=1.05\columnwidth]{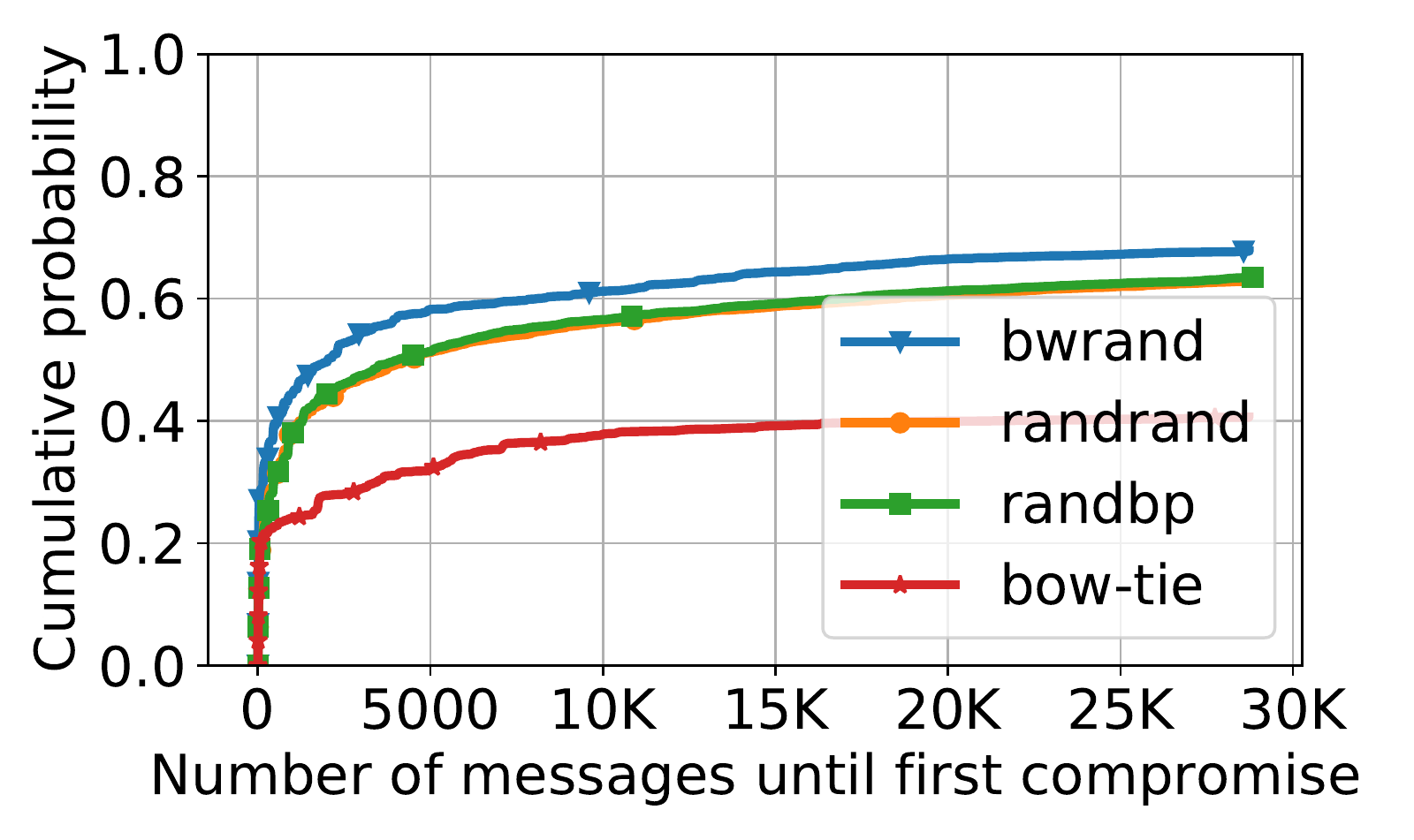}
    \caption{Number of messages until first compromised message.}
		\label{fig:withguard_count}
	\end{subfigure}
  \caption{Enabling client-side guard logic for reference algorithms and comparing the results.}
	\label{fig:dynamic_withguard}
\end{figure}

\subsubsection{Turn on Client Guard-logic for Reference Methods?}
We now turn on the client guard-list logic for \textit{all} designs, \sys and
the references, since the client component is fundamentally independent of the
layer construction algorithm. For the reference 
designs the client will select initial and replacement relays from the middle
layer using the client guard-list logic. 
This will allow us to gauge the effect of the client guard-list on designs without an engineered guard 
layer. Figure~\ref{fig:dynamic_withguard} shows the results of this comparison.  Note that the results we
provide here are independent of the Guard's position in users'
routes.
We see that all the reference designs improve with client guard-logic enabled. However, it is clear that 
\sys enjoys a significantly higher time to first compromise metric than 
the reference designs with client guard-list logic enabled. This means that the guard layer provides an 
added benefit that the client guard-list by itself does not provide, providing at least a $30\%$ 
improvement over the most similar reference design RandBP.

%We also see that the guard layer of Bow-Tie 
%contributes to at least $30\%$ improvements on user's privacy from the gap between Bow-Tie and 
%RandBP.
This confirms the necessity of 
both Bow-Tie's guard layer and client guard-logic that combined reduce clients' \textit{guard exposure} 
more effectively than they each could alone. 

%% file: 6_extension.tex
\section{Influence of Protocols and User Behavior}
\label{sec:extension}

In general, user anonymity is significantly impacted by aspects that we
organize into three broad and independent families: 
topological design choices, Mixnet protocol designs, and user behavior. 

Our discussions and analysis so far concerned 
topological design choices, which refer to engineering aspects of the network itself (such as our guard 
design) to maximize 
users' expected anonymity. Other designs such as Atom~\cite{kwon2017atom} or
XRD~\cite{kwonxrd} add strong topological constrains making the anytrust assumption
realistic and trustworthy, but at the price of severe performance impact limiting potential network 
use-cases and wide adoption. It is up to the user, and or application designer what trade-off is 
appropriate for their use-case.
%making the network use-cases and chances for a consequent user-base limited.

%However, topological engineering is one part of the picture. Protocol integration and user behavior are 
%two other dimensions that 

So far we have not considered the impact of protocol integration or client usage, which, if done 
carelessly, 
may nullify the benefits of Bow-Tie's topological design choices. For 
example, BiTorrent~\cite{inria-00471556} exchanges IP information with a tracker required by
its application-level protocol. For this reason, tunneling BiTorrent inside an anonymous
network does not provide anonymity protection, yet this user activity is observed in
Tor~\cite{bittorent-in-tor}. In the same vein, for Mixnets we cannot tunnel many
existing protocols as-is because it may similarly also nullify the Mixnet's protection. For example, in email,
SMTP and IMAP servers contain many pieces of meta-information that can link users to their activity, and 
even the plaintext if the user does not manually set-up
end-to-end encryption (which requires advanced understanding of threats, email,
and technology). It is also the case for secure messaging applications, such as
Signal, which leverage a central server to enable confidential
communications (while exposing the users' social graph to the central server). To mitigate these threats, 
the Mixnet protocol suite has to offer the means~\cite{dake_unger} to perform
asynchronous messaging, which applications could then use to build secure and private
protocols. 

Similarly, user behavior also has a significant impact. For users sending a single message in
the network, we can evaluate the user's anonymity via entropic considerations.
Different entropy measures may capture different
criteria~\cite{murdoch2008metrics, syverson2009m} and lead to different interpretation of the user's 
anonymity.

We now bring these aspects into our investigation to round out our evaluation of Bow-Tie.

\subsection{Influence of Protocol Designs}
\label{subsec:proto-inf}

We now consider the impact of a recipient anonymity property over the Mixnet
protocol design. Recipient anonymity may be needed to improve users anonymity
in some context, and unnecessary in others. For example, uploading a file to a
public-facing server would not be recipient-anonymous. Exchanging messages
asynchronously with a peer at a private address would be recipient-anonymous.
To obtain recipient-anonymity, we assume the existence of a private and secure
Naming scheme and rendezvous protocol (also called ``dialing'')
defined by the Mix network~\cite{tordesign, van2015vuvuzela, tyagi2017stadium,
  karaoke}.

There is a rich history of anonymous networks that claim strong anonymity~\cite{chaum1981untraceable, 
	chaum1988dining,
	freedman2002tarzan, danezis2003mixminion, sirer2004eluding,
	wolinsky2012dissent, corrigan2015riposte, van2015vuvuzela, kwon2016riffle,
	lazar2016alpenhorn, piotrowska2017loopix, tyagi2017stadium, kwon2017atom, karaoke, KatzMixnet, 
	diaz2021nym}. 
Thoroughly studying and comparing the influence of those various designs is out
of scope of this paper. However, we can explore how simple design choices
can lead to significant recipient anonymity improvements for continuous-time Mix
networks. 

\begin{figure}[!t]
	\centering
    \includegraphics[width=0.25\textwidth]{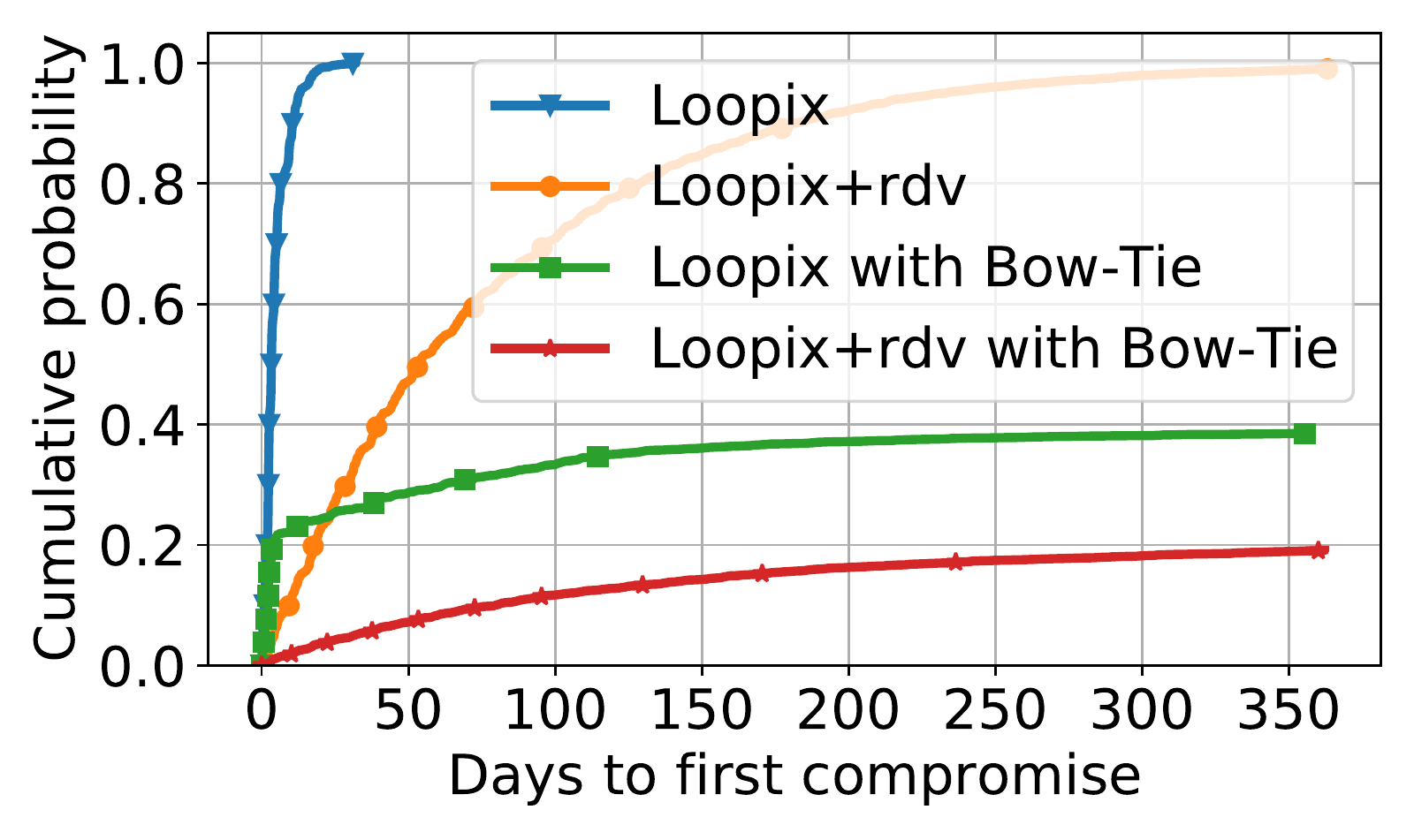}
    \caption{Influences of protocol interaction: benefit of rendezvous-based protocols over
	no-recipient anonymity.} 
	\label{fig:loopix_vs_vuvu_metrics} 
\end{figure}

\begin{figure}[!t]
		\centering
		\includegraphics[width=0.25\textwidth]{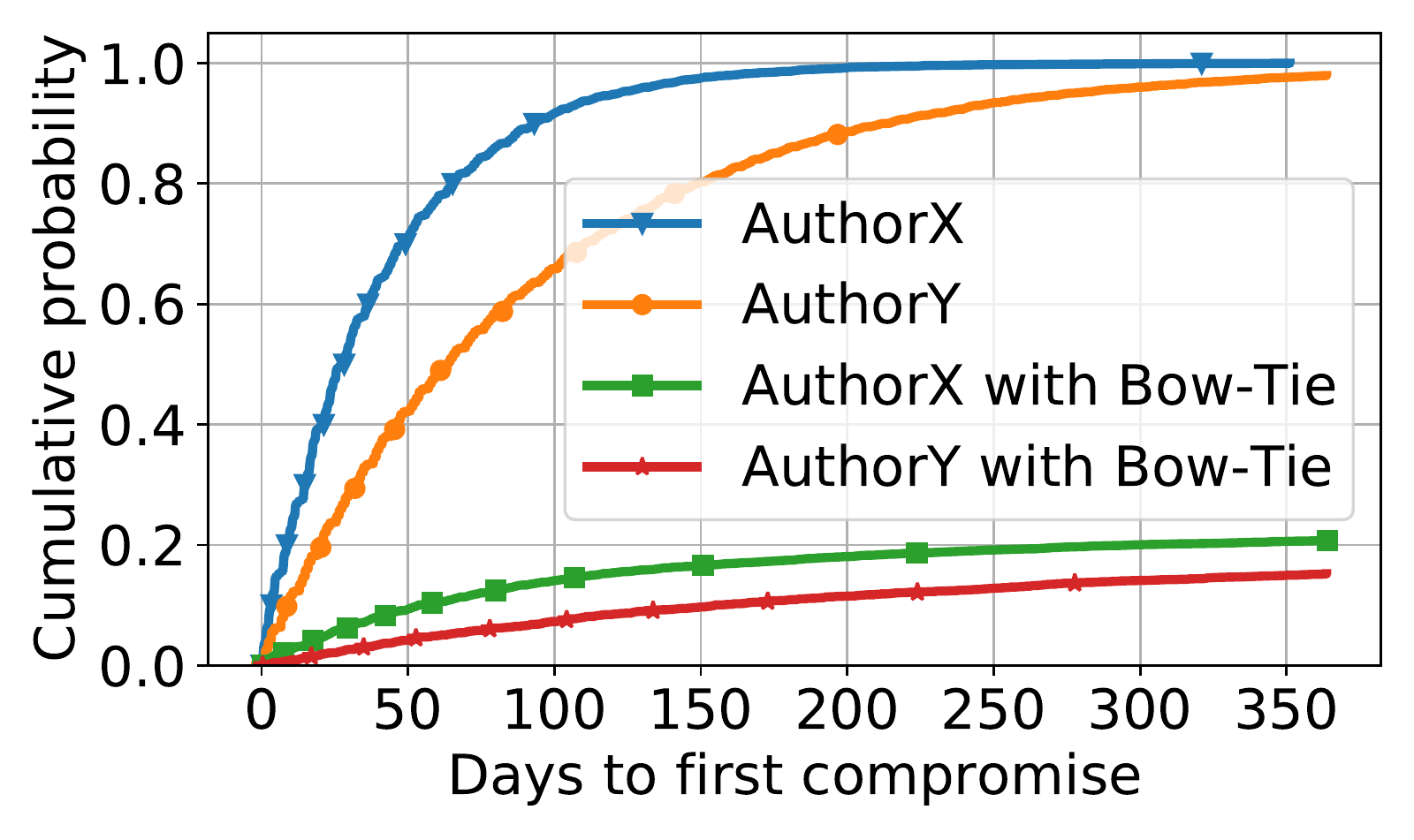}
		 \caption{Influences individual behaviours: comparing the email behavior of two authors with/without Bow-Tie designs.}
		\label{fig:time_email}
\end{figure}

% \begin{figure}[!t]
%     \begin{subfigure}[t]{0.225\textwidth}
%     \centering
%     \includegraphics[width=\textwidth]{figures/loopix_ttfc_cdf_routesimresults.pdf}
%     \caption{Benefit of rendezvous-based protocols over
% 	no-recipient anonymity.} 
% 	\label{fig:loopix_vs_vuvu_metrics} 
% 	\end{subfigure}
% 	\begin{subfigure}[t]{0.225\textwidth}
% 		\centering
% 		\includegraphics[width=\textwidth]{figures/email_ttfc_cdf_routesimresults.pdf}
% 		 \caption{Comparing the email behavior of two authors with/without Bow-Tie designs.}
% 		\label{fig:time_email}
% 	\end{subfigure}
% 	\caption{Influences of protocol interaction and individual behaviours.}
% 	\label{fig:emails_analysis}
% \end{figure}

% \begin{figure}
% 	\centering
% 	\includegraphics[width=0.5\columnwidth]{figures/loopix_ttfc_cdf_routesimresults.pdf}
%   \caption{Benefit of rendezvous-based protocols (e.g., Vuvuzela, Stadium) over
% 	no-recipient anonymity (Loopix, Nym). The user model is based on a
% 	real-world Email sending pattern built from a University staff members' dataset.
%   The network churn rate between each epoch (set to 24h) is 3\% and the number of hops to destination 
%   or rendezvous is 3. \xinshu{2) give the motivation on why the paremeters in this setting were chosen as such: e.g., 1-day epoch.}
% } 
% 	\label{fig:loopix_vs_vuvu_metrics}
% \end{figure}

Loopix~\cite{piotrowska2017loopix} and the Nym Network~\cite{diaz2021nym} based
on the Loopix design do not offer recipient anonymity for asynchronous messages
between clients.  Other designs
such as Tor~\cite{tordesign}, Vuvuzela~\cite{van2015vuvuzela},
Stadium~\cite{tyagi2017stadium} and Karaoke~\cite{karaoke} do offer it thorough a rendezvous 
protocol (also called ``dialing'') to
asynchronously connect peers both seeking to communicate together anonymously. \texttt{routesim} 
can 
model
the two approaches and evaluate the benefit of rendezvous-based protocols,
with respect to users' activity and the path length. The user model is based on a real-world Email sending 
patterns built from a dataset of University staff members. The dataset was 
built from meta-info 
contained within the
\texttt{sendmail} logs from the university SMTP server over a period of two
months with the sending habits of hundreds IT staff members
from the authors' faculty.~\footnote{See Appendix~\ref{app:ethics} for ethics details.}
The network churn rate between each epoch is $3\%$ and the number of hops to destination or rendezvous is $3$.
Figure~\ref{fig:loopix_vs_vuvu_metrics} shows an evaluation of a typical Email
sending pattern derived from the dataset. We see that designs with rendez-vous protocol have better 
security and in combination with \sys are significantly more secure. Given
those results, existing
deployments, such as Nym, may find valuable to incorporate recipient anonymity. However, there is a 
cost to obtaining 
recipient anonymity this way; it doubles the bandwidth consumption for asynchronous messaging, and 
requires
the establishment of an out-of-bound solution to propagate addressing
information. Many different approaches have been detailed in the
literature~\cite{torspec, van2015vuvuzela, lazar2016alpenhorn}. Privately
accessing and retrieving the mailbox contents~\cite{express} is also a
potential approach to gain recipient anonymity, yet would limit the size of the
anonymity set to the number of mailboxes stored on a given mixnode and be
significantly more CPU costly.

%In future work, we intend to explore protocol integration 
%more thoroughly. 

\subsection{Evaluating Individual Risks}
Our earlier analysis considered a simple client. We now consider and
 evaluate complex personal usage behaviors. We built
several datasets containing typical weekly behavior from years of our own
email communication patterns and fed them into \texttt{routesim}. Knowing how
we behave in a typical period of one week, \texttt{routesim} plays a sequence
of events (i.e. sending emails)---that statistically matches our recorded behavior---indefinitely through 
time.
Note that \texttt{routesim} could also simulate other usage patterns, provided a
dataset is available.

In \texttt{routesim}, many configuration options are possible. For this
experiment, we assume each user has a set of ten contacts, use the Bow-Tie
topology with a $3\%$ Mixnode churn rate and an epoch of $1$ day, set
the route length of the Mix network is three, and assume that the Mix network exposes
a protocol suite for asynchronous messaging offering anonymity for both
communicants (i.e. a naming scheme and a rendezvous protocol). 
The Mix network carries the same quantity of data as was typically
contained in the authors' sending email patterns, rounded up to a product of
the Mixnet message payload length ($2048$ bytes).
Essentially, a sender
sends the (end-to-end encrypted) message to the recipient's Mailbox located
within one of the Mixnodes. The recipient anonymously retrieves the Mailbox
contents on demand. The protocol to check and retrieve the
encrypted content is assumed to be derived from a PIR protocol~\cite{chor1995private} to
avoid leaking which Mailbox is queried to the Mixnode. In \texttt{routesim}, we
assume (it is configurable) that a user's Mailbox changes its location at each
epoch (i.e., handled by a different Mixnode selected at random in the first $N-1$
layers). We advise Mixnet developers to never store any encrypted content
on a Mixnode that can exit to the clearnet, hence to never store Mailboxes on a Mixnode that
can be placed in the $Nth$ layer. 

% \begin{figure} \centering
% 	\begin{subfigure}[b]{0.225\textwidth}
% 		\centering
% 		\includegraphics[width=1.05\textwidth]{figures/email_ttfc_cdf_routesimresults.pdf}
% 		\label{fig:time_email}
% 	\end{subfigure}
% 	\begin{subfigure}[b]{0.225\textwidth}
% 		\centering
% 		\includegraphics[width=1.05\textwidth]{figures/email_counts_emails_cdf_routesimsresults.pdf}
% 		\label{fig:counts_email}
% % 		\caption{}
% 	\end{subfigure}
%   \caption{Comparing the email behavior of two authors over Bow-Tie
%     topologies, while disabling and enabling the guard design. The
%     Mix network carries the same quantity of data than what was typically
%     contained in the authors' sending email pattern, rounded up to a product of
%     the Mixnet message payload length (2048 bytes).\xinshu{Still has the problem of unfinished lines. I couldn't fix it.}}
% 	\label{fig:emails_analysis}
% \end{figure}

Figure~\ref{fig:time_email} shows the time to compromise the first pair\footnote{The authors 
are 
always the message senders in these pairs.} of
communicants with email-like communication patterns.  The simulated users
\texttt{AuthorX} and \texttt{AuthorY} have a different email-like communication
pattern in terms of frequency leading to a significant difference in the time to
first compromise. 
%#######comment out for now, as figure right is weird
% Observe however that the metrics ``Number of emails until
% first compromise'', Figure~\ref{fig:emails_analysis} (right), shows a reverse situation, in which AuthorY 
% now seems to
% suffer more. This is a direct consequence of what we could call the \emph{average
% exposure} of the user emails to Mixnodes. Indeed, at any given time, we need
% the author's contacts to be exposed to a malicious Mailbox to suffer potential
% deanonymization. 
In this simulation, users change their mailbox location every
day, meaning that all emails sent the same day to the same contact are all
guaranteed to be exposed to the same Mailbox. Therefore, with this design
choice, the more the user's emails are sparsely sent in time, the more likely
they are exposed to different contact's Mailboxes. Different design choices
would lead to different results. For example, users could decide to change
their Mailbox location not on a day-by-day basis, but rather dependent on the
number of email messages which they fetched. Eventually such design choice
needs to be enforced by the Mixnet developers, and can be evaluated with
\texttt{routesim}.

In the same vein, given an established design such as Bow-Tie, end users such as journalists or
whistleblowers can use \texttt{routesim} to evaluate their chance of being
deanonymized assuming a realistic mix adversary. Results obtained may help them
evaluating whether the risks are worth their information.

%% file: 6_related_work.tex
\section{Related Work}
\label{sec:related_work}
%atom
The literature is rich of Mixnets proposals~\cite{chaum1981untraceable, chaum1988dining,
freedman2002tarzan, danezis2003mixminion, sirer2004eluding,
wolinsky2012dissent, corrigan2015riposte, van2015vuvuzela, kwon2016riffle,
199325,
lazar2016alpenhorn, piotrowska2017loopix, tyagi2017stadium, kwon2017atom, KatzMixnet,
kwonxrd, diaz2021nym, express, spectrum}. Many of them put forward an anytrust
assumption over the network routes to address the insider threat. Some of these
proposals, such as Atom~\cite{kwon2017atom} or XRD~\cite{kwonxrd}, discuss how to shape the network
for this anytrust assumption to be more realistic. These discussions relate to
our approach. However, they are applied to radically different Mixnet designs
than the continuous-time mix design~\cite{danezis2004traffic,
  piotrowska2017loopix, nym2021}.  In our work, we shape the network to
minimize the adversarial impact. We observe that the research effort presented
here is necessary to the practicability of those proposed designs.

Some works investigate the detection and mitigation of active malicious mixes 
and combine it with the Mixnet construction design. Dingledine and Syverson
~\cite{dingledine2002reliable} discuss how to build a mix cascade network 
through a reputation system that decrements the reputation score of all nodes 
in a failed cascade, and increments the reputation of nodes in a successful 
cascade. 
% With the communal random seed generated by all nodes, the system will 
% choose nodes for each cascade randomly from a pool of mixes that have highest 
% reputation values. 
It improves the reliability of mixnet and reduce the chance 
that an adversary controls an entire cascade. However, some pitfalls are 
introduced by the reputation system and the actual deployment is still a 
complex problem. Leibowitz et al. propose Miranda~\cite{
leibowitz2019no}, another reputation-based design that detects and isolates active 
malicious mixes. They also discuss how to construct the cascade mixnet based 
on their faulty mixes detection scheme and a set of cascades are selected 
randomly for the upcoming epoch. This design relies on a fixed set of mixes 
and it is still challenging to deploy in the real world.

Nym~\cite{nym2021} network is designed to support privacy-enhanced access to 
applications and services with metadata being protected based on Loopix~\cite{piotrowska2017loopix}.
The stratified network is periodically constructed from a large number of available mixes 
run by profit-motivated mix operators, who are compensated for their investment 
with payment in Nym's cryptocurrency tokens. Nym's design~\cite{diaz2021nym},
is sketched out in a whitepaper, 
presenting their solution to construct a Mixnet by 
% lacking technical depth, of how
% and the network is constructed 
% by 
randomly selecting
mixes weighted by mixes' stake and randomly placing them 
into layers. 
Nym uses a verifiable random function 
(VRF)~\cite{Micali1999VerifiableRF} to facilitate the features of decentralization in their blockchain-based ecosystem. In our work, we expect
that sampling from bandwidth is a good proxy to the process of sampling from
stake and we abstract Nym's sampling as a sampling by bandwidth,
with meaningful bandwidth values borrowed from the Tor network (instead of
% <<<<<<< HEAD
ad-hoc stake values) as BwRand (Section~\ref{subsec:reference_algos}).
% =======
% ad-hoc stake values) as BwRand.
% >>>>>>> 90b31c4807784e30d9820ed508d20c9425a0d9fe

% In Nym, the stratified network is periodlically (re)constructed from a random
% subset of the available mixes run by profit-motivated mix operators. These
% operators are compensated for their investment with payment in Nym's
% cryptocurrency tokens. The relevant detail is that some fraction of all
% available mixes are selected for inclusion in the network based on their
% stake, i.e. third-parties' trust in the ability of the relay to perform
% according to its claimed capacity. In our terminology, we call this subset the
% \textit{active set} of mixes. In the Nym design, adversarial manipulation of
% the constructed Mixnet is mitigated by using a verifiable random function
% (VRF)~\cite{Micali1999VerifiableRF} to place the mixes into a desireable
% stratified topology. Paths through the network are baised by relay bandwidth
% at each hop of the route.

% With a large number of available mix nodes, only a subset is selected to route
% traffic in the network with a probability that is proportional to the mixes'
% stake. We call this set of nodes the \emph{active set}. All stakeholders in
% Nym can put stakes (which acts as money) into any node to increase its chance
% of being selected as part of the network. Then it randomize the placement of
% nodes via a verifiable random function (VRF) ~\cite{Micali1999VerifiableRF},
% in order to prevent the adversary from manipulating the assignment of
% malicious mixes in the network.

Guirat and Diaz~\cite{guirat1mixnet} investigate how
to optimize the Mixnet parameters for a continuous-time mix
network, and focus on
the number of layers and the width of the network (i.e., the number of nodes in
each layer). They theoretically analyze the fully compromised rate for a
continuous-time mix network in a designated shape and they mainly concentrate
on optimizing the Mixnet parameters using the Shannon entropy~\cite{
serjantov2002towards, serjantov2003anonymity, diaz2002towards} as the 
guiding metric. 
% In contrast, we focus on the
% problem of how to limit the adversary's end-to-end compromising ability by
% modeling a realistic mix network and strategic adversary, and thouroughly investigate how to construct 
% the network, and how to select path that led to the \sys design.

%The paper also consider dummy strategies to provide good mixing against a
%global passive adversaries.
%They mainly focus on the specific kind of Mixnets with same number of malicious
%nodes and honest nodes in each layer and calculate the fully compromised rate
%(i.e., worst case scenario) as one of the security metric. They compare the
%difference in this metric between two path selection methods by studying a
%specific shape of Mixnet: in a $3 \times 10$ mixnet where among $10$ mixes in
%each layer there is $1$ malicious mix node, uniform path selection is set as
%messages routing by equal-bandwidth mix nodes, while the bandwidth of malicious
%node is set to $40\%$ of the total layer capacity in bandwidth-weighted path
%selection scenario. They conclude that the adversary gains less advantage from
%populating malicious mixes in the case of uniform path selection. Based on
%this, they study, with uniform routing, the optimal value of Mixnets
%parameters.

%% file: 7_conclusion.tex
\section{Conclusion}
\label{sec:conclusion}

In this paper, we address the question of "how to shape the Mixnet to strengthen the 
anytrust assumption?" and study the design of Mixnet configuration and routing that 
limits the adversary's power to deanonymize traffic. We proposed Bow-Tie, 
a practical and efficient novel design for mix network engineering; we present the first thorough security analysis of stratified Mixnet against reasonably realistic adversaries;
we develop the \textit{routesim} simulator that can easily calculate users' expected 
deanonymized probability. In the future, we will further explore the case with untrusted
configuration server.

% \xinshu{add future work for the case of untrusted configuration server, using weighted 
% verifiable selection and verifiable placement.}

%% file: 8_appendix.tex
\section{Guard's Position Considerations}
\label{app:choosing_position}

In Section~\ref{subsec:design_overview}, we present the Guard idea for
Continuous-time mixnets which aims at reducing users' exposure to malicious
mixnodes. Choosing the position of the Guard in users' path of length $L$ is an
interesting question leading the following analysis:

\begin{itemize}
  \item Choosing the last layer could allow a malicious guard to perform
    re-identification attacks based on prior knowledge. For example, if the
    network is used to connect to user-dependent destinations (Services,
    set of contacts), then the \textit{a priori} knowledge of this relation would reveal the identity of the 
    mixnetwork user.
  \item Choosing a layer in  $[2..L-1]$ has the advantage, compared to Tor, to
    not directly bind the long-lived guard to the user. That is, discovering
    the identity of a user's guard does not lead directly to the user,
    i.e. the Guard's ISP can not be compelled to reveal the client IP addresses connecting to the guard 
    relay.
    Low-latency anonymity networks such as Tor cannot move the guard's position
    into some layer $[2..L-1]$ as their threat's model expect end-to-end
    traffic confirmation to succeed in deanomymizing a user-destination
    relation. Therefore, for a low-latency design, moving the guard to a layer
    $[2..L-1]$ would achieve nothing. We do not have this issue with
    Continuous-time mix-networks.
  \item Choosing the first layer has a massive performance advantage in
    continuous-time mix networks using Sphinx packets~\cite{danezis2009sphinx}, the state
    of the art packet format specification for Continuous-time mix networks.
    Indeed, currently, a full cryptographic handshake is performed for each
    Sphinx packet, which is needlessly costly when all packets are sent to the
    same first node (the guard), and only one cryptographic handshake for a
    determined session period would lead to much lower performance impact. 
    
    One possible method is for clients to perform
    L-1 Sphinx processing (for hops $2..L$) and $1$ TLS processing for the first
    hop. We do a small scale experiment for preliminary indicative results. We compare the throughput of a
    Rust sphinx implementation~\cite{sphinx-implem} with a \texttt{AES-128-GCM}
    openssl benchmark, the most used cipher in TLS1.3, over $1024$ bytes blocks.
    The choice of $1024$ bytes comes from the default Sphinx packet size choice.
    Over a AMD Ryzen 7 3700X, we were able to perform 8261 Sphinx unwrap/s for a
    payload of 1024 bytes. With \texttt{AES-128-GCM}, we processed $\approx
    500\times$ more packets per seconds. Moreover, TLS has a maximum payload size
    of 16KiB, which means that multiple sphinx packets can be encrypted within the
    same record, leading to a performance improvement of $\approx 600\times$, on
    average from our benchmarks.
\end{itemize}

Therefore, choosing the guard layer position is a trade-off between user anonymity and performance. 
Users'
anonymity benefits from guards in layer $[2..L-1]$, while network
performance benefits from guard position in the first layer, and a
enhanced first packet processing design described next.

Our experimental analysis and results in this paper are independent of the
Guard's position within user's path. We leave choice of trade-off to the implementer.

\section{Email Dataset Ethics}
\label{app:ethics}
The University email dataset collection ethics application was filed with the faculty's ethics process 
(application \#41564). This was approved prior to the IT department initating the collection of the data. 
Only select meta-data (from email headers) was collected relevant to email sending patterns. All personal 
information in the headers was pseudonymized before we were given access.

\section{Empirical Results of Random Path Selection}
\label{sec:random path selection}

We evaluate the security and performance of Bow-Tie and reference methods with random
path selection using the metric of \emph{compromised fraction of paths} and \emph{expected queuing delay}. Note that the adversary's best resource allocation policy under random 
path selection is to inject as many malicious Mixnodes as possible, since the quantity matters more than bandwidth. 
% the best adversarial \textit{even resource allocation policy} for 
% compromising high traffic with random path selection is to inject as many 
% nodes as possible into the network, with the minimal allowed bandwidth. 
In our simulation, we 
% utilize this for the 
% random path selection (\ref{subsubsec:compromised traffic}), and 
instantiate this 
best strategy as generating thousands of Mixnodes with a minimum of $1$\,MBps, since there 
could be an infinite number of malicious Mixnodes if we do not set a lower 
bound.

% in
% the case of random path selection (top half), the compromised rates are high when there
% are a large number of injected malicious Mixnodes

\begin{figure}[th]
	\centering
	\begin{subfigure}[t]{.49\columnwidth}
		\centering
		\includegraphics[width=\textwidth]{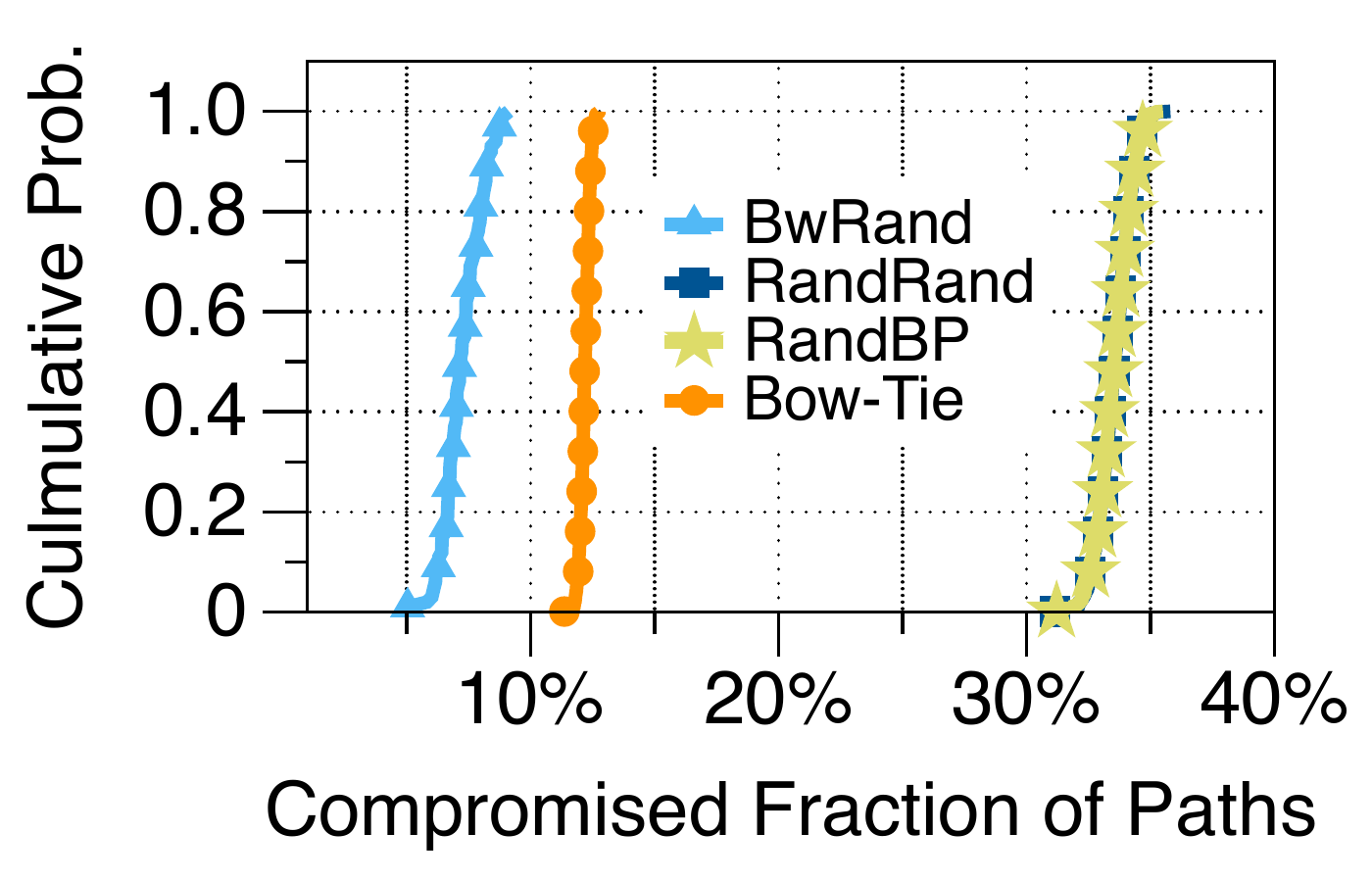}
		\caption{Probability distribution on compromised fraction of paths with $h=0.75$.}
		\label{subfig:best_rand_075}
	\end{subfigure}
    \begin{subfigure}[t]{.49\columnwidth}
        \centering
        \includegraphics[width=\textwidth]{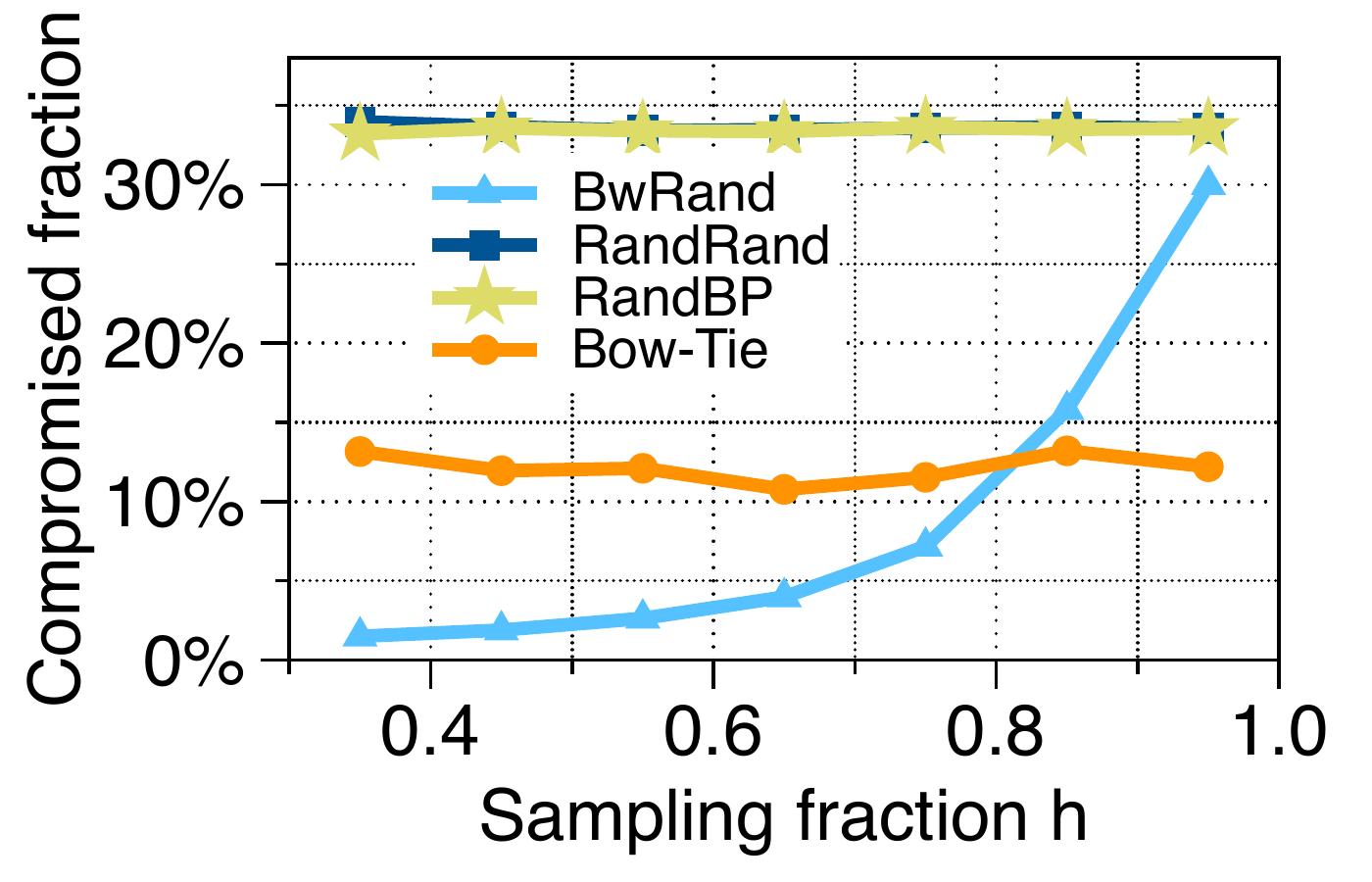}
        \caption{Average compromised fraction of paths for $h$ between $0.35$ to $0.95$.}
        \label{subfig:rand_all_a02}
    \end{subfigure}%
	\caption{Compromised fraction of paths using random path selection, $\alpha=0.2$. }
	\label{fig:security_rand}
\end{figure}

\begin{inparaenum}
\item \emph{Security evaluation.} We set the compromised fraction of paths $F_r$ in a 
stratified $l$-layer Mixnet using random message forwarding as

\begin{equation}
\label{equ:fully comp rate random}
F_r = \prod_{i=1}^{l} \frac{\text{Number of Malicious Mixnodes in Layer $i$}}{\text{Number of Mixes in 
Layer $i$}}.
\end{equation}

Figure~\ref{subfig:best_rand_075} shows that, when $h=0.75$, BwRand limits the 
compromise rate between $5\%$ and $9\%$ with relatively higher security guarantee in 
comparison to other methods. 
% In the scenario of random path selection, Figure~\ref{subfig:rand_all_a02} shows the
% simulated results of average compromised traffic and
% Figure~\ref{subfig:best_rand_035},~\ref{subfig:best_rand_055},~\ref{subfig:best_rand_075}
% show the CDFs of the empirical results. 
% RandRand and RandBP are the most
% vulnerable in random path selection where the adversary could de-anonymize much
% more traffic than his resource proportion.
% , while they provide the highest
% protection against adversarial manipulation in the scenario of
% bandwidth-weighted path selection. 
By looking at Figure~\ref{subfig:rand_all_a02}, we also see that 
BwRand mitigates the adversary's
compromising power in a wide range of $h$ and provides the best protect in this case. 
Therefore, BwRand coupled
with a uniform path selection may appear to be an interesting candidate.
However, as shown in Figure~\ref{subfig:rand_all_a02}, the best compromise rate that we can get 
from BwRand\&Random path selection (RPS) is around $1.89\%$ with $h=0.35$, which is comparable to 
the worst compromise rate that we obtain from Bow-Tie\&Bandwidth-weighted path selection (BPS) is around 1.92\% with $h=0.35$ (Figure~\ref{subfig:bw_all_a02}). Besides, BwRand\&RPS shows a dramatic increase as $h$ increase while Bow-Tie\&BPS enjoys a stable security level.

\item \emph{Performance evaluation.}
Suppose there are $n$ nodes in one layer, then 
the expected queuing time in random path selection setting for this 
layer is:
\begin{equation}
\label{equ:rand_expect_time}
T_r = \sum\limits_{i=1}^{n}\frac{n^{-1} u_i 
(2-n^{-1}u_i\Lambda)}{2(1-n^{-1}u_i\Lambda)}.
\end{equation}

\begin{figure}[th]
\centering
\includegraphics[width=0.24\textwidth]{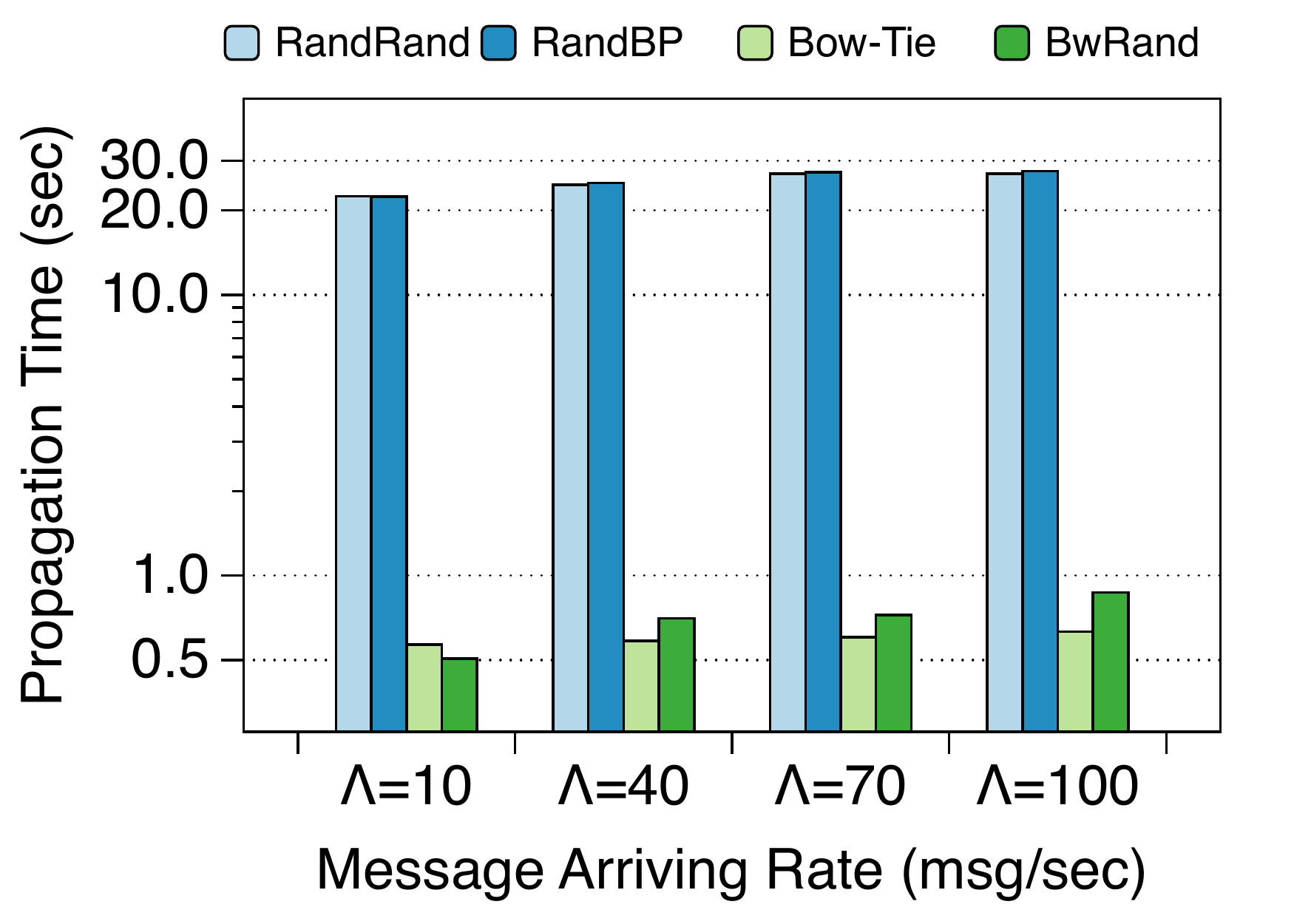}
\caption{Expected queuing delay, with message arriving rate $\Lambda$ for the Mixnet 
based on random path selection.}
\label{fig:rand_queue}
\end{figure}
\end{inparaenum}

Figure~\ref{fig:rand_queue} shows the expected delay due to queuing for a message 
going through the 
Mixnet with random path seleciton. 
Still, algorithms that sample using bandwidth
(i.e., BwRand and Bow-Tie) achieve relatively low processing delay and outperform random sampling 
schemes. 
However, the Mixnet takes more time to handle 
handles one order of magnitude low message
arrival rates than in Bandwidth-weighted path selection.

\section{Adversary Resource Allocation}
\label{app:resource allocation}
\begin{figure}[th]
\centering
\includegraphics[width=0.3\textwidth]{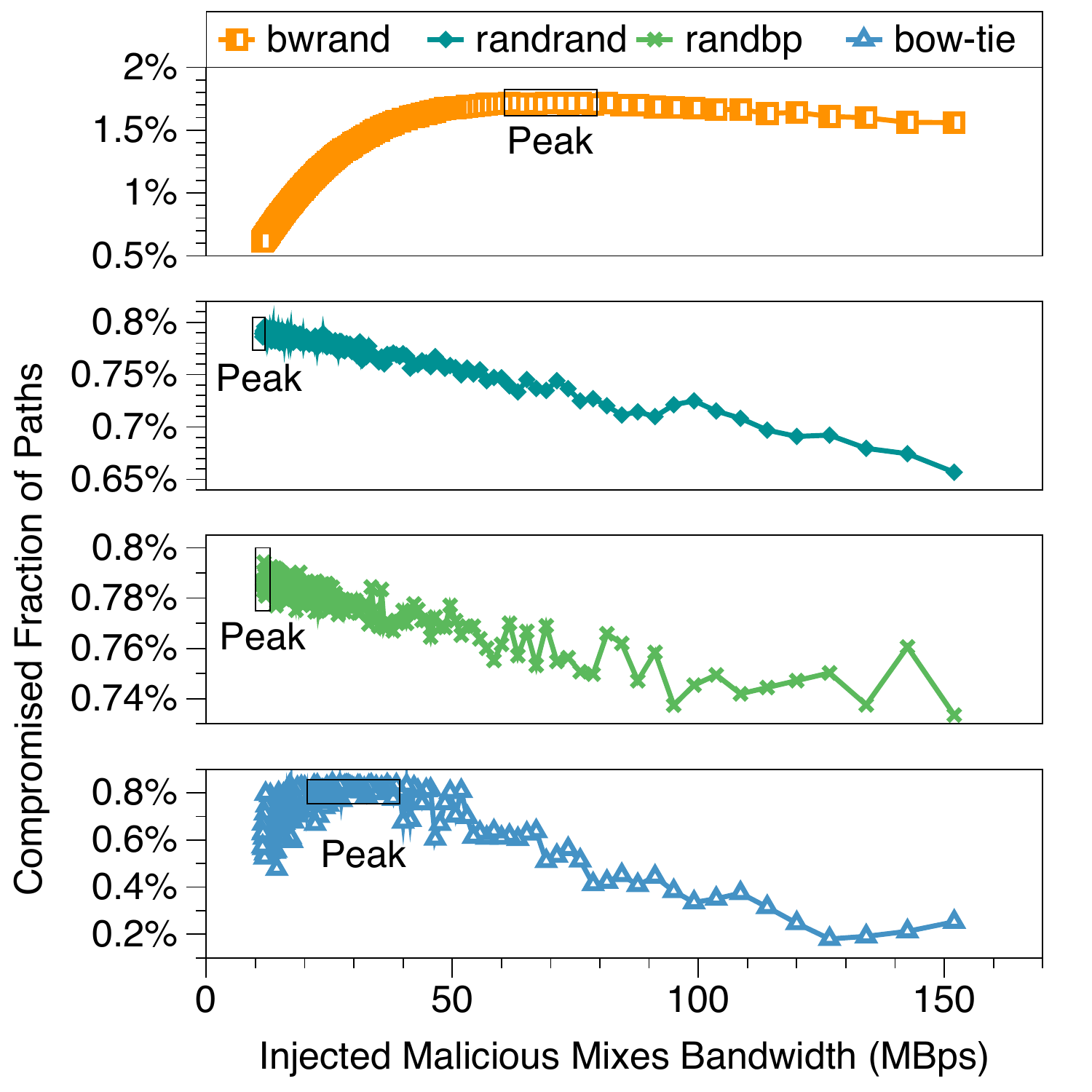}
\caption{Fully compromised fraction versus bandwidth per injected malicious node. The adversary controls $2280$ MBps bandwidth which is allocated to a number of equal-size Mixnodes.}
\label{fig:four_resource_allocation}
\end{figure}

The results, displayed in Figure~\ref{fig:four_resource_allocation}, show that the compromised fraction of the adversary (Section~\ref{subsec:security metrics}) for different algorithms. The optimal capacity sizes of Mixnodes that maximizes the compromising rate are aligned with 
the information shown in Figure~\ref{fig:layer_shape} and also confirms our choices of best resource allocation strategy (in Section~\ref{subsubsec:security_methodology}).

\newpage
\section{Algorithms}
\label{app:algorithms}
    \begin{algorithm}[t!]
    \DontPrintSemicolon
  % \SetNoFillComment
    \SetKwProg{func}{Function}{:}{end}
    % \SetKwFunction{Init}{Init}
    \SetKwFunction{MaintainGuardSet}{MaintainGuardSet}
    \SetKwFunction{BSSample}{BSSample}
    \SetKwFunction{IBSSample}{IBSSample}
    \SetKwFunction{TotalBw}{TotalBw}
    \small
    \KwIn{candidate mix pool $P$ with $P_{bw}$ bandwidth; sampling fraction $h$; tolerance fraction $\tau$.}
    \KwOut{configured guard layer $L_g$ for upcoming epoch $i$; updated guard set $G$.}
    \BlankLine
    \If(\tcp*[f]{Initialize the guard layer}){first call}{
        % \Return \Init{$P$, $h$, $\tau$}
        Sample $\frac{h}{3} * P_{bw}$ nodes from $P$, weighted by bandwidth, as a set $AG$\;
        Sample $\tau * \frac{h}{3} * P_{bw}$ nodes from $P-{AG}$, weighted by bandwidth, as a set $BG$\;
      % $L_g\longleftarrow$ Sample $\frac{h}{3} * \TotalBw(P)$ from $P$, weighted by bandwidth\;
      % ${BG}\,\longleftarrow$ Sample $\tau * \frac{h}{3} * \TotalBw(P)$ from $P\,-\,{L_g}$, weighted by bandwidth\;
      % Give nodes in $AG$ and $BG$ a common label $G$\;
      $G\,\longleftarrow\,{AG}\,\cup\,{BG}$\tcp*[r]{Give nodes in $AG$ and $BG$ a common label $G$}
      Place all ${AG}$ nodes into guard layer $L_g$\;
      \ForEach{$ag$ in $AG$}{
      $t_{AG}\,\longleftarrow\,1$\tcp*[f]{Track working time as a guard}
      }
      \Return $L_g$, $G$\;
      }
    \Else(\tcp*[f]{Maintain the guard layer}){
    % $AG', BG' \longleftarrow AG, BG$\;
    % $AG, BG \longleftarrow \emptyset, \emptyset$\;
    Update Mixnodes on/off status\;
    Update Mixnodes stability metric (\textit{WMTBF})\;
    % $\MaintainGuardSet{$G$}\tcp*[r]{Update Guard Set}
    $G,\,{DG}\,\longleftarrow\,$\MaintainGuardSet{$G$}\;
    \ForEach{$g$ in $G-{DG}$}{
    \If(\tcp*{Inherit old online $ag$}){$t_{AG} > 0$}{
    Move node $g$ to $AG$\;
    }
    }
    $BG\,\longleftarrow\,G\,-{DG}\,-\,{AG}$\; 
    % $DG \longleftarrow $ Offline nodes in $G$\;
    % $L_g \longleftarrow \ G-DG$\tcp*[r]{Copy old AG that are still online}
    % ${AG} \longleftarrow $ nodes in $G-DG$ with $AGT > 0$\;
    $\delta \longleftarrow$ \TotalBw($AG$) $-$ $T_{low}$\;
    \If(\tcp*[f]{Insufficient $ag$}){$\delta < 0$}{
    ${AG}$ ${+=}$ \BSSample{${BG}$, $\left|\delta\right|$}\tcp*[r]{Add $\left|\delta\right|$
    nodes from $BG$}
    % ${roG}\,\longleftarrow\,G\,-\,{DG}\,-\,{AG}$
    % ${AG}\,\longleftarrow\,{AG}\,\cup\,$\BSSample{${roG}$, $\left|\delta\right|$}
    }
    Place all ${AG}$ nodes into guard layer $L_g$\;
    \ForEach{$ag$ in $AG$}{
    update $t_{AG}$\tcp*[l]{Track working time as a guard}
    % $t_{AG}\,{+=}\,1$\tcp*[f]{Track working time as a guard}
    }
    % $L_g\,\longleftarrow\,{AG}$\;
    % $BG\,\longleftarrow\,G\,-\,{DG}\,-\,L_g$\;
    \Return $L_g$, $G$\;
    }
    % \func{\Init{$P$, $h$}, $\tau$}{
    %   Sample $\frac{h}{3} * \TotalBw(P)$ nodes from $P$, weighted by bandwidth, and flag them as $AG$\;
    %   Sample $\tau * \frac{h}{3} * \TotalBw(P)$ nodes from $P-{AG}$, weighted by bandwidth, and flag them as $BG$\;
    %   % $L_g\longleftarrow$ Sample $\frac{h}{3} * \TotalBw(P)$ from $P$, weighted by bandwidth\;
    %   % ${BG}\,\longleftarrow$ Sample $\tau * \frac{h}{3} * \TotalBw(P)$ from $P\,-\,{L_g}$, weighted by bandwidth\;
    %   Place all ${AG}$ nodes into $L_g$\;
    %   \ForEach(\tcp*[f]{Track working time as a guard}){$ag$ in $AG$}{
    %   $t_{AG}\,\longleftarrow\,1$\;
    %   }
    %   $G \longleftarrow AG \cup BG$\tcp*[r]{Label all nodes as Guards}
    %   \Return $L_g$, $G$\;
    % }
    \func{\MaintainGuardSet{$G$}}{
    Gather offline nodes in $G$ to a subset $DG$\;
    % $DG\,\longleftarrow\,$offline nodes in $G$\;
    $\delta_l \longleftarrow$ \TotalBw{$G-DG$} $-$ $T_{low}$\;
    $\delta_h \longleftarrow$ \TotalBw{$G-DG$} $-$ $T_{high}$\;
    \uIf(\tcp*[h]{Too few online guards.}){$\delta_l < 0$}{
    $G\,{+=}\,$\BSSample{${P-G}$, $\min{\{ \left|\delta_l\right|,\,\TotalBw{${P-G}$} \}}$}\;

    % ${newG}$ $\longleftarrow $ \BSSample{$P-G$, $\left|\delta_l\right|$}\;
    % $G \longleftarrow G\,\cup\,{newG}$\;
    }
    \uElseIf(\tcp*[h]{Too many online guards}){$\delta_h > 0$}{
    % rmv$G$ $\longleftarrow $ \IBSSample{nodes in G with $t_AG=0$, $\left|\delta_h\right|$}\;
    % $\delta_h' \longleftarrow$ \TotalBw{$G-DG$} $-$ upper bound of $L_g$\;
    % \If{\TotalBw{$DG$} $<$ $\left|\delta_h\right|$}{
    $S\,\longleftarrow$ \{$g$ with $t_{AG}=0$\}\;
    % \tcp*[r]{Gather online fresh $g$ as to a set $S$}
    $G\,{-=}\,$\IBSSample{$S$, $\min{\{ \left|\delta_h\right|,\,\TotalBw{S} \}}$}\;
    % 
    % }
    % \Else{
    % $G\,{-=}\,$\IBSSample{$DG$, $\left|\delta_h\right|$}\;
    % }
    }
    \Return $G, DG$\;
    }
    
    \func{\BSSample{$T$, $k$}}{
    Normalize \textit{WMTBF} to $0-1$ scale for nodes in $T$\;
    Sort all nodes by \textit{bw}$\times$\textit{WMTBF} in descending order\;
    % \If{\TotalBw{$T$} $\ge k$}{
    % $\min{\{ k,\,\TotalBw{T} \}}$\;
    $S \longleftarrow$ Mixnodes that add up to $\min{\{ k,\,\TotalBw{T} \}}$ bandwidth in order\;
    % }
    % \Else{
    % $S\,\longleftarrow\,T$;
    % }
    $T\,{-=}\,S$\;
    \Return $S$\;
    }
    \textbf{Note:} function \IBSSample{} is the same as \BSSample{} except sorting all nodes 
    in an inverse order.
    % \func{\IBSSample{$T$, $k$}}{
    % % Normalize \textit{WMTBF} to $0-1$ scale for nodes in $T$\;
    % Sort all nodes by \textit{bw}$\times$\textit{WMTBF} in ascending order\;
    % % $S \longleftarrow$ select Mixnodes that add up to $k$ bandwidth in order.\;
    % % $T\,{-=}\,S$\;
    % % \Return $S$\;
    % }
\caption{Configuring Guard Layer}
\label{alg:config_guard}
\end{algorithm}

      \begin{algorithm}[t]
      \small
      \DontPrintSemicolon
      \SetKwFunction{TotalBw}{TotalBw}
      \KwIn{candidate mix pool excludes guard nodes $P' = P - G$; sampling fraction $h$.}
      \KwOut{Configured two layers $L_l, L_r$ for upcoming epoch $i$.}
      \BlankLine
      $L_l, L_r \longleftarrow \emptyset$\;
      $P_{Active} \longleftarrow$ sample $\frac{2}{3}h * P_{bw}$ Mixnodes uniformly from $P'$\;
      % \ForAll(\tcp*[f]{Binpacking placement}){$mix \in P_{Active}$}{
      $n \longleftarrow |P_{Active}|$\tcp*[f]{Binpacking placement starts}\;
      $W \longleftarrow \emptyset$\;
      \For(\tcp*[f]{prepare weights for ILP}){$j \longleftarrow0$ \KwTo n}{
      $W \longleftarrow W \cup b_j$\;
      }
      $c \longleftarrow (\frac{h}{3} + \epsilon) * P_{bw}$\tcp*[r]{expected capacity for each layer}
      $l \longleftarrow 2$\tcp*[r]{number of projected layers}
      $L_l, L_r \longleftarrow ILP(W, l, c)$\;
      \Return $L_l, L_r$\;
      % }
    \caption{\small Configuring Non-guard Layers}
    \label{alg:config_ordinary}
    \end{algorithm}